\documentclass[twocolumn]{aa}  

\usepackage{natbib,twoopt}
\usepackage[breaklinks=true]{hyperref} 
\bibpunct{(}{)}{;}{a}{}{,}             
\makeatletter
  \newcommandtwoopt{\citeads}[3][][]{\href{http://adsabs.harvard.edu/abs/#3}%
    {\def\hyper@linkstart##1##2{}%
     \let\hyper@linkend\@empty\citealp[#1][#2]{#3}}}
  \newcommandtwoopt{\citepads}[3][][]{\href{http://adsabs.harvard.edu/abs/#3}%
    {\def\hyper@linkstart##1##2{}%
     \let\hyper@linkend\@empty\citep[#1][#2]{#3}}}
  \newcommandtwoopt{\citetads}[3][][]{\href{http://adsabs.harvard.edu/abs/#3}%
    {\def\hyper@linkstart##1##2{}%
     \let\hyper@linkend\@empty\citet[#1][#2]{#3}}}
  \newcommandtwoopt{\citeyearads}[3][][]%
    {\href{http://adsabs.harvard.edu/abs/#3}
    {\def\hyper@linkstart##1##2{}%
     \let\hyper@linkend\@empty\citeyear[#1][#2]{#3}}}
\makeatother

\usepackage{gensymb}
\usepackage{graphicx}
\usepackage{txfonts}
\usepackage{gensymb}
\usepackage{multirow}
\usepackage{xspace}
\usepackage{amsmath}

\newcommand{\kms}{km$\,$s$^{-1}$\xspace}
\newcommand{\nnh}{N$_2$H$^+$\xspace}
\newcommand{\nnd}{N$_2$D$^+$\xspace}

\newcommand{\dco}{DCO$^+$\xspace}
\newcommand{\hco}{HCO$^+$\xspace}
\newcommand{\hcdo}{HC$^{18}$O$^+$\xspace}
\newcommand{\tex}{$T_\mathrm{ex}$\xspace}
\newcommand{\tmb}{$T_\mathrm{MB}$\xspace}
\newcommand{\ncol}{$N_\mathrm{col}$\xspace}

\begin{document}

   \title{High sensitivity maps of molecular ions in L1544: I. Deuteration of \nnh and \hco  and first evidence of \nnd depletion. \thanks{This work is based on observations carried out with the IRAM 30 m Telescope. IRAM is supported by INSU/CNRS (France), MPG (Germany) and IGN (Spain).}}


   \author{E. Redaelli
          \inst{1}
          \and
         L. Bizzocchi\inst{1} \and
         P. Caselli \inst{1} \and
         O. Sipil\"a \inst{1} \and
         V. Lattanzi \inst{1} \and
         B. M. Giuliano \inst{1} \and
         S. Spezzano \inst{1}}
           
  \institute{Centre for Astrochemical Studies, Max-Planck-Institut f\"ur extraterrestrische
              Physik, Gie\ss enbachstra\ss e 1, D-85749 Garching bei M\"unchen (Germany) \\
              \email{eredaelli@mpe.mpg.de}
               }

     \titlerunning{Deuteration maps in L1544}
     
   \authorrunning{Redaelli et al.}
   \date{Received ****; accepted *****}

 
  \abstract
   {The deuterium fraction in low-mass prestellar cores is a good diagnostic indicator of the initial phases of star formations, and it is also a fundamental quantity to infer the ionisation degree in these objects.}
   {With the analysis of multiple transitions of \nnh, \nnd, \hcdo and \dco we are able to determine the molecular column density maps and the deuterium fraction in \nnh and \hco toward the prototypical prestellar core L1544. This is the preliminary step to derive the ionisation degree in the source.}
   {We use a non-local thermodynamic equilibrium (non-LTE) radiative transfer code, combined with the molecular abundances derived from a chemical model, to infer the excitation conditions of all the observed transitions, which allows us to derive reliable maps of each molecule's column density. The ratio between the column density of a deuterated species and its non-deuterated counterpart gives the searched deuteration level.}
   {The non-LTE analysis confirms that, for the analysed molecules, higher-J transitions are characterised by excitation temperatures $\approx 1-2\,$K lower than the lower-J ones. The chemical model that provides the best fit to the observational data predict the depletion of \nnh and to {a lesser} extent of \nnd in the innermost region. The peak values for the deuterium fraction that we find are {$\mathrm{D/H_{N_2H^+}} = 0.26^{+0.15}_{-0.14}$  and $\mathrm{D/H_{HCO^+}} = 0.035^{+0.015}_{-0.012}$, in good agreement with previous estimates in the source.}}
   {}


   \maketitle
%

\section{Introduction}
Deuterium was formed during the primordial nucleosynthesis in the first minutes after the Big Bang with a fractional abundance of $\approx 1.5\times10^{-5}$ \citep{Linsky03}. However, enhancements {of several} orders of magnitude of this value have been found both in different environments of the interstellar medium (ISM) and in several components of the Solar System (\citealt{Ceccarelli14}, and references therein). This has led to the idea of using the deuteration ratio D/H (the ratio between the abundance of a D-bearing molecule and its hydrogenated isotopologue) as a diagnostic tool to investigate the star formation process, and ultimately to understand how our own Solar System was formed \citep{Ceccarelli14}. \par
Prestellar cores represent the early phases of low-mass star formation \citep{Andre14}. These are dense (central volume densities $n \gtrsim 10^5\,\text{cm}^{-3}$) and cold ($T \lesssim 10\,$K) fragments of molecular clouds, which are on the verge of gravitational collapse, but still have not formed any central object {\citep{Keto08}}. Prestellar cores are known to exhibit some of the highest levels of deuteration ($\mathrm{D/H} > 0.1$, \citealt{Crapsi05, Pagani07}), because their physical conditions greatly favour the deuteration processes. D-fractionation (i.e. the inclusion of a D-atom in hydrogenated species), in fact, is driven by the reaction:
\begin{equation}
\label{Deuteration}
\mathrm{H_3^ + + HD} \leftrightharpoons \mathrm{H_2D^+ +H_2}
\end{equation}
which proceeds more efficiently from left to right as the temperature decreases {(assuming a low ortho-to-para H$_2$ ratio; e.g. \citealt{Pagani92})} due to the lower zero-point energy of $\mathrm{H_2D^+}$, the progenitor of all the deuterated species {in regions where the temperature is $\lesssim30\,$K} \citep{Dalgarno84}. Thus, in cold environments, the increased abundance of $\mathrm{H_2D^+}$ and of the doubly and triply-deuterated forms of $\mathrm{H_3^+}$ rapidly brings to higher D/H ratios in all molecules produced by the reactions with the various H$_3^+$ isotopologues. This process is favoured also by another mechanism. The $\mathrm{H_3^+}$ main destruction route is its reaction with CO. Above $ n \approx \text{a few } 10^4\,\text{cm}^{-3}$, CO and other C- and O-bearing species catastrophically freeze onto the dust grain surfaces, leading to their depletion from the gas-phase; in regions where the abundance of neutral species reacting with H$_3^+$ is reduced, D-fractionation is enhanced \citep{Dalgarno84}. In prestellar cores, where the depletion fraction of CO is $>90$\% in the central $6000-8000\,$AU \citep{Caselli99, Bacmann02}, the D-fraction {in $\rm NH_3$} reaches levels as high as 40\% \citep{Crapsi05}. \par
A quantity that plays a crucial role in driving the timescales of star formation is the ionisation degree, i.e. the ratio between the free electron density $n(e)$ and the gas density $n(\mathrm{H_2})$ [$x(e) = n(e)/n(\mathrm{H_2})$] \citep{McKee89}. In fact, this factor determines the time scale for the ambipolar diffusion of neutrals across the magnetic field lines, which is the main process that allows the gravitational collapse of magnetised cores \citep{Mouschovias76}. In molecular clouds, the deuteration fraction is in general considered a good indicator of the ionisation degree \citep{Guelin77,Caselli98, Caselli02b, Dalgarno06}, as the D-fraction in species such as \hco is a function of $x(e)$. As a consequence, accurate measurements of the D/H ratio allow to determine the ionisation degree in prestellar cores. In particular, \nnh and \hco represent ideal tracers \citep{Caselli98, Caselli02c}. These molecules --- and their deuterated counterparts--- are abundant in low-mass cores. \nnh is usually highly concentrated in the central parts, whereas \hco traces the outer layers (e.g. \citealt{Punanova18}). Combined, they can provide information on the ionisation degree in the whole source. 

\par
L1544 is one of the most studied low-mass prestellar cores. Its vicinity ($d \approx 135\,$pc, \citealt{Schlafly14}) and bright continuum emission at mm and sub-mm wavelengths \citep{WardThompson99, Doty04, Spezzano16} make it the ideal environment to study the early stages of low-mass star formation. The central part of the core is cold ($T \approx 6\,$K, \citealt{Crapsi07}) and dense ($n >10^6\,\text{cm}^{-3}$, \citealt{Tafalla02,Keto10}). It is centrally peaked and shows signatures of {contraction \citep{Caselli02a, Lee01, Caselli12}}, suggesting that it is on the verge of gravitational collapse. In the last 20 years, a vast observational campaign combined with theoretical efforts has led to a deep knowledge of its physical and chemical structure (see for instance \citealt{Crapsi07, Caselli12, Keto15, Spezzano17}). Previous works \citep{Caselli98,Caselli02b} have already investigated the deuteration level and ionisation fraction in this source, using the telescope capabilities available at that time.  We have now access to new, high sensitivity maps of \nnh, \nnd, \hcdo, and \dco across the whole core performed with the On-The-Fly method. For the first time, we can analyse at the same time the first three rotational transitions of \nnd and \dco and the \nnh (1-0) and (3-2) lines\footnote{\nnh (2-1) transition is difficult to observe {(although not impossible, \citealt{Daniel07}),} because it falls in the vicinity of a strong atmospheric line.}. In this paper we report the deuteration fraction maps of \nnh and \hco in L1544, obtained with a detailed non-LTE modelling of the molecular spectra at the core's centre. In a following work (Paper II) we will use the results presented here to accurately derive the ionisation fraction map of L1544.\par
In Section \ref{Obs} we present the observations. The analysis is described in Sec. \ref{Analysis}, divided in chemical modelling (Sec. \ref{ChemModSec}), non-LTE modelling (Sec. \ref{non-ltemod}) and column density derivation (Sec. \ref{ColDens}). We discuss the results in Sec. \ref{Discuss}, and draw a summary of the work in Sec. \ref{Conclusions}.

\section{Observations\label{Obs}}
All the data presented in this paper were collected with the IRAM 30m single dish telescope. We used the four receivers of the frontend EMIR (E0, E1, E2 and E3), which allow {us to cover} the required frequency range. The \nnh (1-0) data were acquired in summer 2015. The observations of \nnh (3-2), \nnd (1-0), (2-1) and (3-2), \hcdo (1-0), \dco (1-0), (2-1) and (3-2) were performed during winter 2015/2016 using four different spectral setups. The VESPA backend, with a spectral resolution of $20\,$kHz, was used for all data except for the \dco (1-0) transition, which was acquired with the FTS backend (resolution: $50\,$kHz). In all the sessions, the pointing was frequently checked on nearby planets or bright quasars, and found to be accurate within $2''$ and $5''$ (at the high and low frequencies, respectively). In order to encompass the whole core, all the transitions were acquired in the On-The-Fly (OTF) map mode with a $2' \times 2'$ ($\approx 0.08 \times 0.08\, \text{pc}^2$) footprint centred at the L1544 dust peak (R.A.(J2000)$ = 05^h04'17''.21$, Dec.(J2000)$ = 25^\text{\degree}10'42''.8$, \citealt{WardThompson99}). The \nnh (1-0) data, observed in a different session, consist of a $4' \times 4'$ ($\approx 0.16 \times 0.16 \, \text{pc}^2$) map across the dust peak.  \par
The spectra, originally calibrated in antenna temperature ($T^*_{\mathrm{A}}$), were converted to main beam temperature ($T_{\mathrm{MB}}$) using the tabulated values for the 30m forward efficiency ($F_{\mathrm{eff}}$) and main beam efficiency ($B_\mathrm{eff}$).  The main features of the observed lines, such as their frequency, angular resolution ($\theta_\mathrm{beam}$) and spectral resolution ($\Delta V_\mathrm{ch}$), are summarised in Table 1. \par
In addition to these data, we also make use of a single pointing observation of the \hco (1-0) line performed with the FCRAO 14m telescope towards the dust peak of L1544. These data have a spectral resolution of $0.07\,$\kms and an angular resolution of $55''$, and were previously published in \cite{Tafalla98}. We refer to their paper for further details.

\begin{table*}[h]
\renewcommand{\arraystretch}{1.4}
\centering
\caption{Main properties of the lines observed at the IRAM 30m telescope. }
\begin{tabular}{cr@{.}lcccr@{.}lc}
\hline
Line                       & \multicolumn{2}{c}{$\nu$ (GHz)\tablefootmark{a}} & $\Delta V_\mathrm{ch}$ (km$\,$s$^{-1}$) & $\theta_{\mathrm{beam}}$ ($''$) & $F_{\mathrm{eff}}/B_{\mathrm{eff}}$ & \multicolumn{2}{c}{Int. time (h)\tablefootmark{b}} & Ref.\tablefootmark{c}\\
\hline
\nnh (1-0)  & 93&173          &  0.06         & 27                            & 0.95/0.81          &   2&8              &      1            \\
\nnh (3-2)  & 279&512         & 0.02           & 11                            & 0.87/0.49         & 11&3         &     1          \\
\nnd (1-0)  & 77&109          &      0.08    & 34                            & 0.95/0.82             &  7&5     & 2 \\
\nnd (2-1)  & 154&217         &     0.04      & 16                            & 0.93/0.72             &  6&7 &   2    \\
\nnd (3-2)  & 231&322         &      0.03      & 12                            & 0.92/0.59              &  9&4     &  2  \\
\dco (1-0)  & 72&039          &         0.20      & 36                            & 0.95/0.82           &   7&1      &    3      \\
\dco (2-1)  & 144&077         &     0.04      & 17                            & 0.93/0.73           &  6&6    &    3      \\
\dco (3-2)  & 216&113         &      0.03        & 12                            & 0.94/0.63         & 2&8       &   3    \\
\hcdo (1-0) & 85&162          &      0.07       & 31                            & 0.95/0.81          &      10&2    &  4       \\
\hline
\end{tabular}
\tablefoot{
\tablefoottext{a}{{The references for the rest frequencies are found in Table \ref{SpecTable} {in Appendix \ref{SpectroscopicConstants}. } }}\\
\tablefoottext{b}{Total observing time.} \\
\tablefoottext{c}{{References used for the collisional coefficients (see Sec. \ref{non-ltemod}):} [1] \cite{Lique15}; [2] scaled from \cite{Lique15}; [3] \cite{Pagani12}; [4] scaled from \cite{Flower99}.}}
\end{table*}

\subsection{Results}
\begin{figure*}
\centering
\includegraphics[width = 0.8\textwidth]{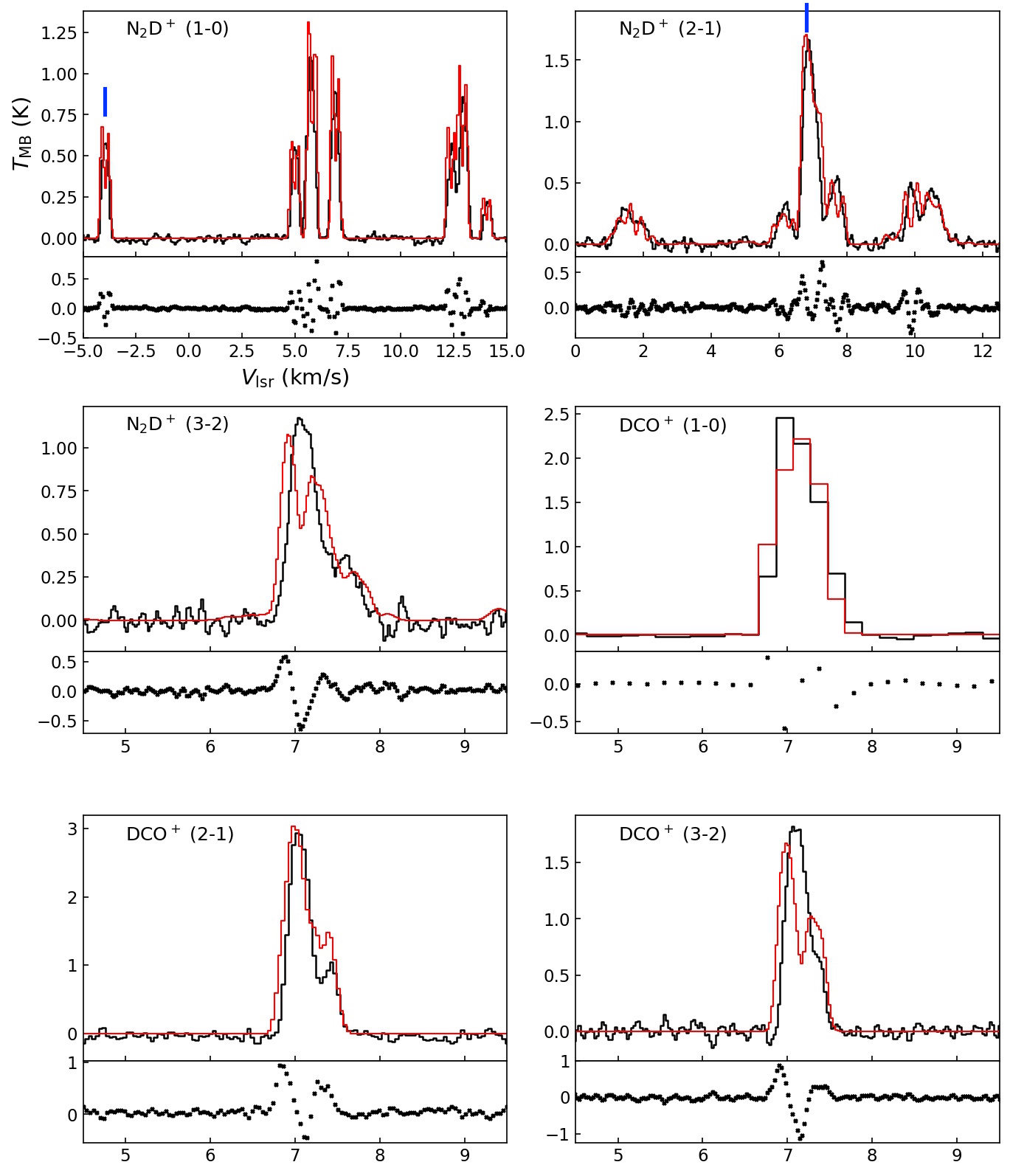}
\caption{{Observed spectra at the dust peak of \nnd and \dco, with the original angular resolution (in black histograms)}. The transition is labeled in the upper-left corner of each panel. In red, the best-fit solution found with MOLLIE are overlaid to the observations (see Sec. \ref{non-ltemod}). {Underneath each line we present the residuals of the fit (difference between the model and the observation).}  The vertical blue bars show which hyperfine component has been used to derive the molecular column density, when the hyperfine structure has not been neglected.  \label{Spectra1}}
\end{figure*}

\begin{figure*}
\centering
\includegraphics[width = 0.8\textwidth]{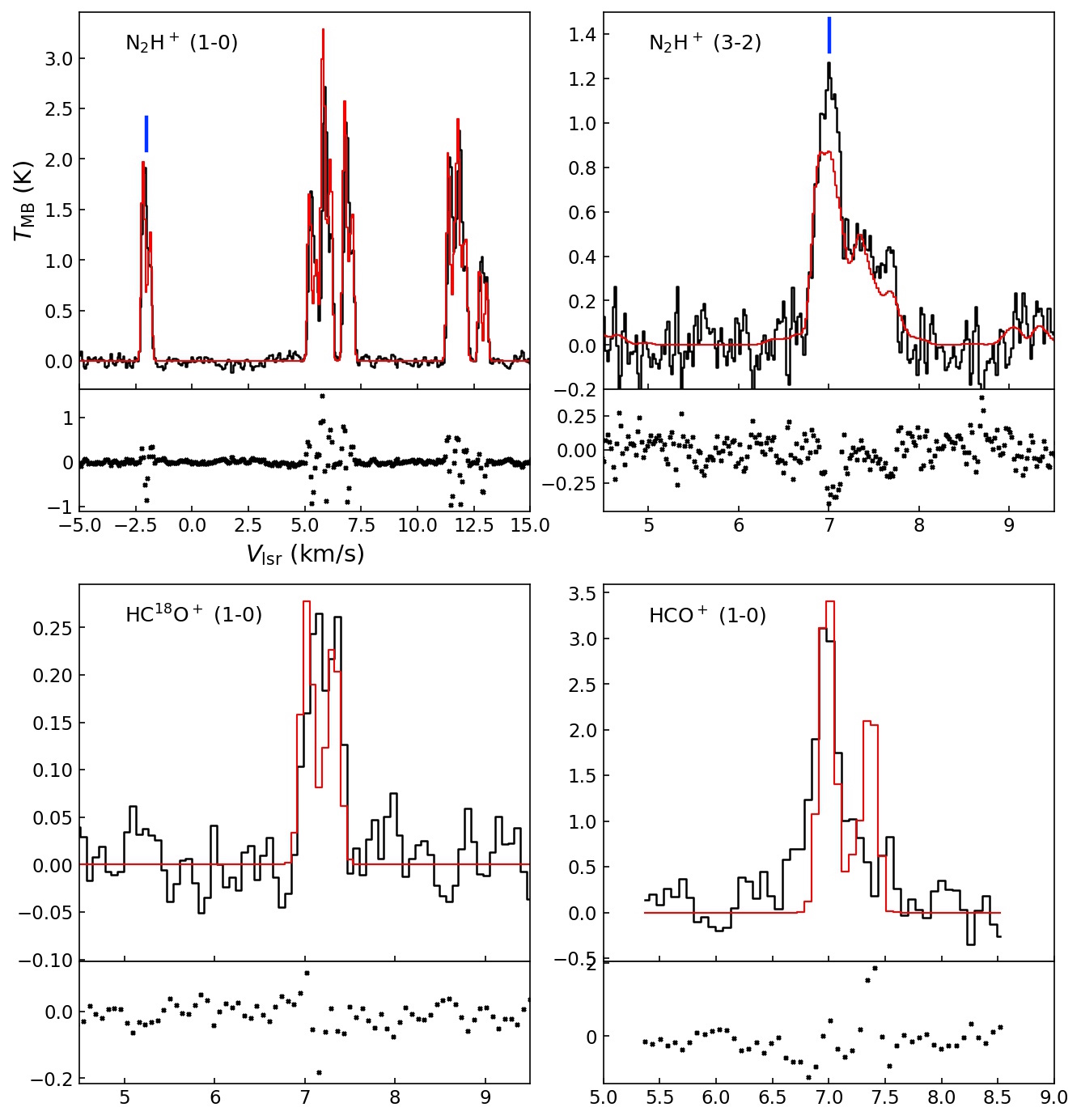}
\caption{{As in Figure \ref{Spectra1}, but for \nnh, \hcdo and \hco transitions. \label{Spectra2}}}
\end{figure*}

\begin{figure*}
\centering
\includegraphics[width = 0.8\textwidth]{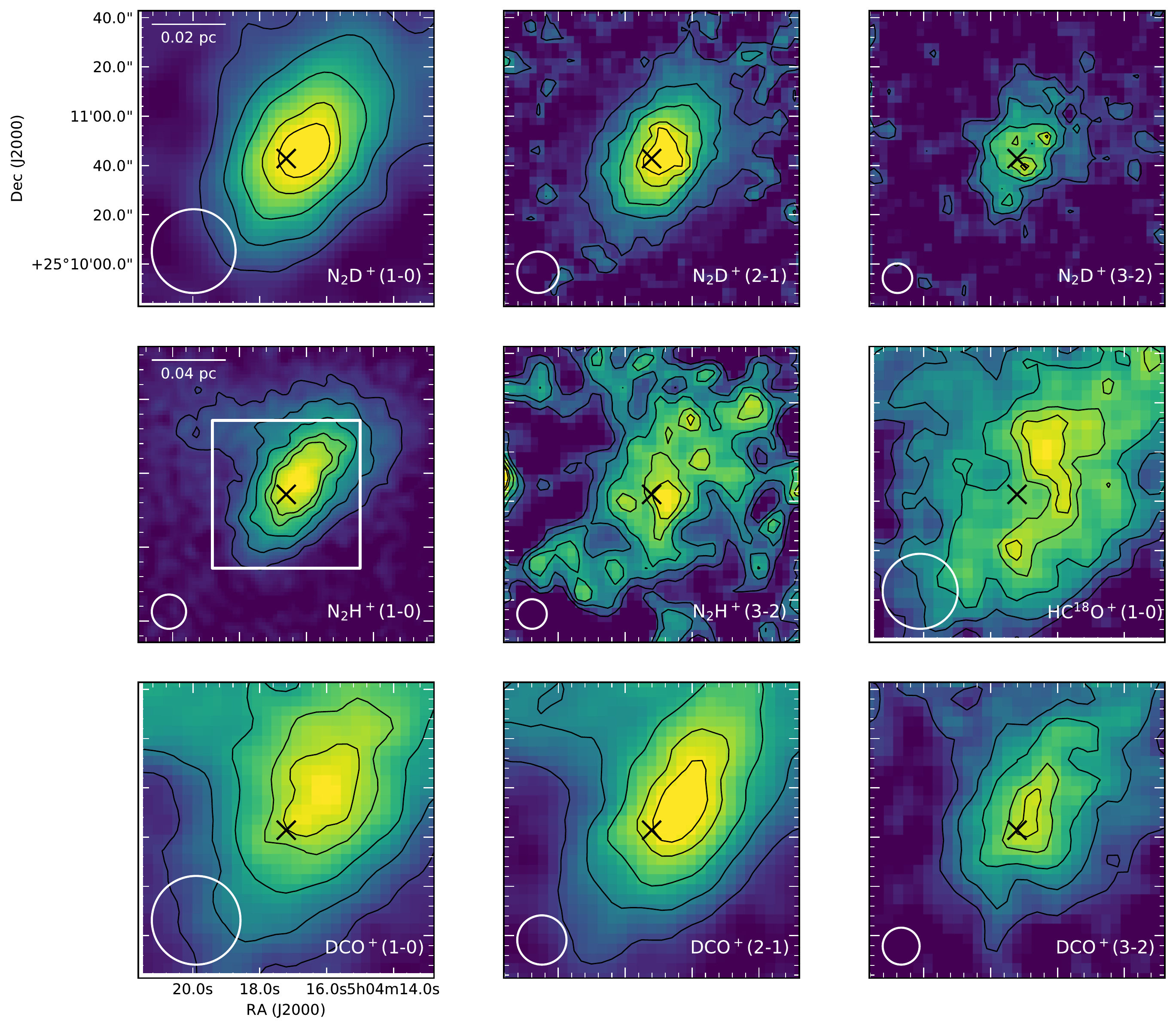}
\caption{Integrated intensity maps of all the observed transitions, which are labeled in the bottom-right corners. The contours show the 20, 40, 60, 80 and 90\% of the peak values, which are in order from top-left to bottom-right panel: 2.0, 2.0, 0.7, 4.6, 0.7, 0.15, 1.8, 1.1, 0.7 K \kms. The black cross represents the dust peak position, and the white circle is the beam size. All the maps have the same size ($2' \times 2'$), with the exception of \nnh (1-0), which is $4'\times 4'$: in this map the white rectangle shows the smaller coverage of the other transitions. {Scalebars are shown in the top-left corners of \nnd (1-0) and \nnh (1-0) maps.} \label{IntInt}}
\end{figure*}

{Figures \ref{Spectra1} and \ref{Spectra2} show the observed spectra at the millimetre dust peak for each transition (in black)}. Considering only the emission-free channels, we computed the mean rms per channel for each transition, obtaining: $0.05\,$K for \nnd (1-0), $0.18\,$K for \nnd (2-1), $0.26\,$K for \nnd (3-2), $0.14\,$K for \nnh (1-0), $0.32\,$K for \nnh (3-2), $0.04\,$K for \hcdo (1-0), $0.08\,$K for \dco (1-0), $0.07\,$K for \dco (2-1), and $0.23\,$K for \dco (3-2). Thus, at the dust peak the observed spectra have a signal-to-noise ratio (SNR) ranging from $\approx 4.0$ to $\approx 40.0$.
Figure \ref{IntInt} shows the integrated intensity maps for each transition, computed integrating over all the hyperfine components when present.


\section{Analysis\label{Analysis}}
Our main goal is to derive the deuterium fraction across the source, and for this aim we need reliable measurements of the column density for each isotopologue. In principle, the knowledge of multiple transitions of a species could allow to determine both its column density and excitation condition (i.e. the line excitation temperature, $T_\mathrm{ex}$). This approach, implemented in the population diagram method \citep{Goldsmith99}, has been extensively used in the literature, but it is valid only in the assumption that all the lines share the same $T_\mathrm{ex}$. This assumptions extends to all the hyperfine components, when present. However, we know from previous works \citep{Daniel06, Daniel13} that this is often not the case for the molecules considered here. {This can be} due to multiple reasons, such as selective photon trapping in very crowded hyperfine structure, or a significant difference in critical density for the different rotational transitions of the same molecule (the Einstein coefficient $A$ scales with the third power of the frequency). {In this case, the column density and the excitation temperatures of the different transitions are degenerate parameters, and they cannot be constrained simultaneously.} Therefore, an approach that does not assume {LTE} is needed to determine each transition's \tex and thus the molecular column density.  \par
A full, non-LTE analysis is on the other hand not feasible on maps, because it requires the knowledge of the full three-dimensional temperature and density structure of the core. For L1544, however, \cite{Keto15} developed a one-dimensional physical model based on a collapsing Bonnor-Ebert sphere {that} provides a good fit to several molecules at the dust peak \citep{Caselli12, Bizzocchi13, Caselli17, Redaelli18}.  We thus decided to model the spectra at the core's dust peak with a full non-LTE analysis. {In this way, we are able to constrain the column {density} of each molecule at the dust peak. At the same time, we can use the equations of radiative transfer to derive $N_\mathrm{col}$ in the whole map. Since these two approaches must return the same result at the core's centre, the \tex (which is involved in the radiative transfer equations) is treated as a free parameter to be tuned in order to obtain the expected agreement of the two methods. The non-LTE modelling allows us to break the degeneracy between $N_\mathrm{col}$ and \tex mentioned in the previous paragraph. In the following Subsections we provide a detailed description of the needed steps.}

\subsection{Chemical models\label{ChemModSec}}
The non-LTE modelling of line emission requires the knowledge of molecular abundance {($X_\mathrm{mol}$)} profiles. To derive these, we used the gas-grain chemical model discussed in \citet{Sipila15a,Sipila15b}, which has recently received an extensive update {\citep{Sipila19}}. Briefly, we employ the static physical core model from \cite{Keto15}, divided into concentric shells, and solve the chemical evolution separately in each shell. We then combine the results at different time steps to produce radius-dependent abundance profiles for the molecules of interest. {The adopted initial conditions are summarised in Table \ref{InitialCond}. More information on typical model parameters can be found for instance in \citet{Sipila15a}.}
\par
In this work we considered five different time steps: $t = 10^4 \, \rm yr$, $t = 5 \times 10^4 \, \rm yr$, $t = 10^5 \, \rm yr$, $t = 5 \times 10^5 \, \rm yr$, and $t = 10^6 \, \rm yr$. We tested the influence of external UV radiation by considering three different visual extinction values for the cloud embedding the core: $A_{\rm V} = 1, 2$, or 5\,mag. This makes it a total of 15 modelled abundance profiles per molecule. {We show the complete set of profiles in Appendix \ref{CompleteModels}}. Simulated lines were produced based on these profiles as described in the following {Subsection}.

\begin{table}
\renewcommand{\arraystretch}{1.4}
	\centering
	\caption{{Initial abundances (with respect to the total hydrogen number density $n_{\rm H}$) used in the chemical modelling. The initial $\rm H_2$ ortho/para ratio is $1.0 \times 10^{-3}$. \label{InitialCond}}}
	\begin{tabular}{cc}
		\hline
		Species & Abundance\\
		\hline
		$\rm H_2$ & $5.00\times10^{-1}$\\
		$\rm He$ & $9.00\times10^{-2}$\\
		$\rm HD$ & $1.60\times10^{-5}$\\
		$\rm C^+$ & $1.20\times10^{-4}$\\
		$\rm N$ & $7.60\times10^{-5}$\\
		$\rm O$ & $2.56\times10^{-4}$\\
		$\rm S^+$ & $8.00\times10^{-8}$\\
		$\rm Si^+$ & $8.00\times10^{-9}$\\
		$\rm Na^+$ & $2.00\times10^{-9}$\\
		$\rm Mg^+$ & $7.00\times10^{-9}$\\
		$\rm Fe^+$ & $3.00\times10^{-9}$\\
		$\rm P^+$ & $2.00\times10^{-10}$\\
		$\rm Cl^+$ & $1.00\times10^{-9}$\\
		\hline
	\end{tabular}
\end{table}

\subsection{Non-LTE modeling at the dust peak \label{non-ltemod}}
The non-LTE approach was implemented using the radiative transfer code MOLLIE \citep{Keto90,Keto04}, which allows us to derive synthetic spectra from a given physical model. MOLLIE is able to treat the line overlap that occurs in a crowded hyperfine structure, such as those of \nnh and \nnd. {As an output, it also provides the molecular column density, obtained integrating the abundance profile multiplied by the gas density, eventually convolved to a desired beam size.} \par
\begin{figure}[h]
\centering
\includegraphics[width = 0.5 \textwidth]{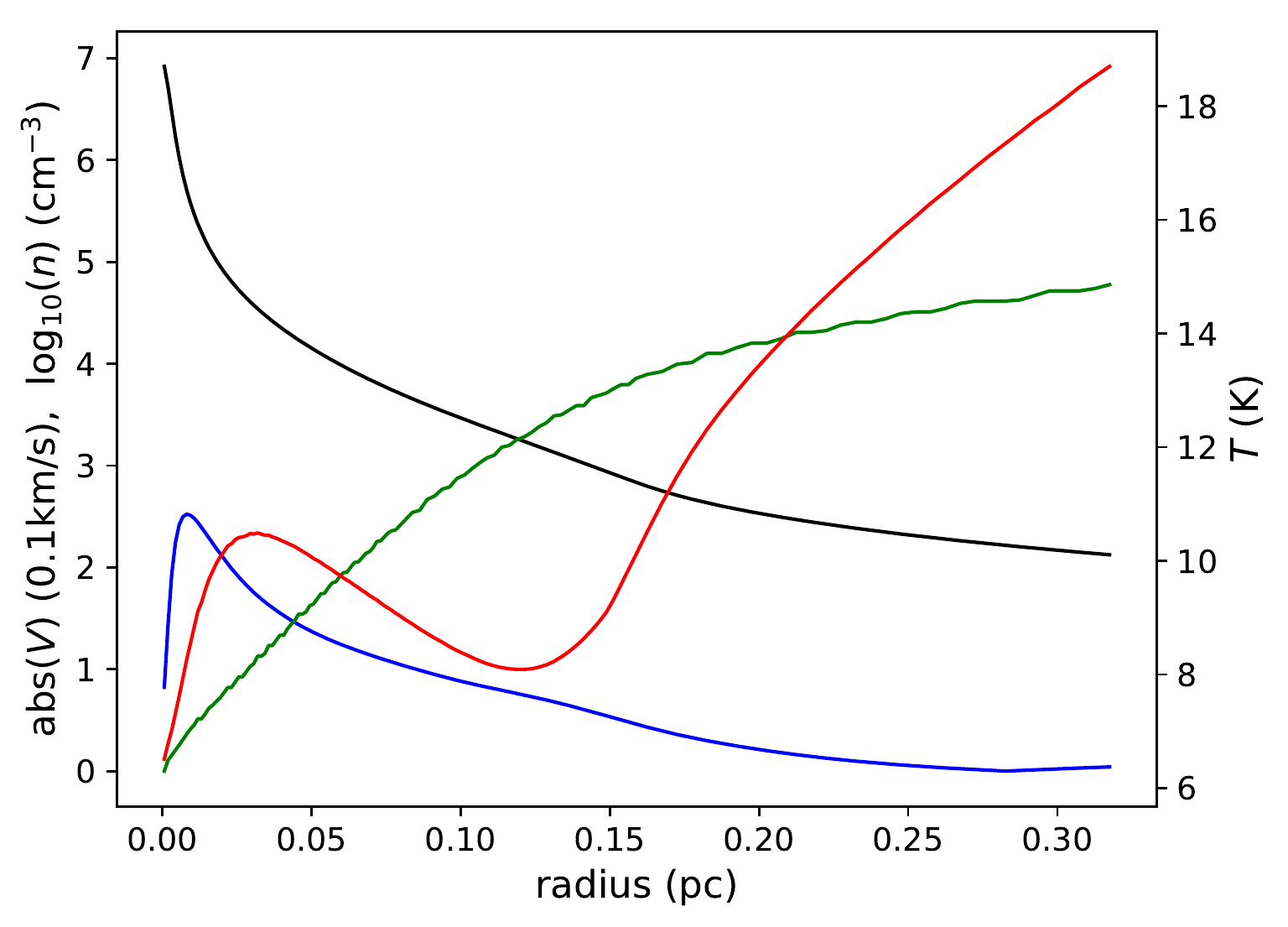}
\caption{{The profiles of dust temperature (green), gas temperature (red), H$_2$ volume density (black, in logarithmic scale), and velocity (blue, in units of $0.1\,$\kms) for the L1544 model developed in \cite{Keto15}. The velocity in the model is negative (it represents contraction motions), but it is shown here as positive for better readability.  \label{KetoModel}}}
\end{figure}

The physical model of the source, as already mentioned, is taken from \cite{Keto15}. It is an unstable quasi-equilibrium Bonnor-Ebert sphere, with a central density $n_0 \approx 10^7\, \mathrm{cm^{-3}}$ and a central temperature of $6.5\,$K and $6.3\,$K for the gas and the dust, respectively. {Figure \ref{KetoModel} shows the profiles of the main quantities for this model.} Recently, \cite{Chacon-Tanarro19} proposed a new density and (dust) temperature model for L1544, analysing new millimetre continuum observations of L1544.  However, the authors did not develop a new model for the velocity field, a crucial quantity to correctly reproduce molecular spectra. Therefore, we prefer to use the previous model, which also includes the velocity 
profile. \par

For the line radiative transfer, the collisional coefficients are of great importance, and we used the most recent ones. When they are available only for one isotopologue, we scaled them taking into account the different reduced mass of the collision system. The collisional coefficients for \nnh are taken from \cite{Lique15}, who computed them for the \nnh/p-H$_2$ system. The rates for the \dco/p-H$_2$ system were computed by \cite{Pagani12}, who take into account the hyperfine structure given by the deuterium nucleus. Finally, the collisions of \hco with p-H$_2$ were published by \cite{Flower99}, from whom we scaled the coefficients for \hcdo. In all cases, the o-H$_2$ is neglected, which is a reasonable assumption given that the ortho-to-para ratio (OPR) in L1544 is found to be as low as $\mathrm{OPR} = 10^{-3}$ \citep{Kong15}. \par
We tested all the 15 models described in Sec. \ref{ChemMod}. The abundance profiles produced by the chemical model often overestimate or underestimate systematically the observed fluxes, and need to be multiplied by a numerical factor to find a good agreement with the observations. However, MOLLIE requires too long computational times to allow a full sampling of the parameter space. For each molecule, we thus proceeded as follows: first, we test all the original models, varying the evolutionary stage and the thickness of the surrounding cloud, and we select the one that best reproduces the trend of the different transitions (for instance, the ratio between the peak temperatures of the brightest component of each line).  Then, we multiply the whole profile by a numerical factor, testing $\approx 5$ values until the observed fluxes are matched. This is considered the best fit to the observations. {We want to highlight that our goal is to obtain a simultaneous fit for all molecules that is roughly consistent with the observations, and a full parameter-space search for the best possible fit within the model uncertainties is beyond the scope of the present paper.} In the following text, we focus on the abundance profiles that provide the best fit solution for each molecule. As an example, we present in { Sec. \ref{n2dpAna}} the resulting spectra from several of the 15 abundance profiles for \nnd.

\subsubsection{Best fit solution for the observed molecules at the dust peak} The best fit {solutions} for each molecule are shown in red in {Figures \ref{Spectra1} and \ref{Spectra2}}, overlaid to the observations. We also include here the analysis of a single-pointing observation of \hco (1-0) performed at the dust peak (see Sec. \ref{Obs} for details). Despite the fact that a map for this transition is not available, it is important to check the behaviour of the chemical model for \hco, the main isotopologue of its family. Its observed spectrum, together with the best fit from MOLLIE, is shown in {the bottom-right panel of Figure \ref{Spectra2}.}

\paragraph{\nnh}
The chemical network systematically underestimates the emission of \nnh, for any combination of chemical time and external visual extinction. We find that in general the spectra are best reproduced at late evolutionary times ($10^{6}\,$yr), when the molecules at the core's centre start to be affected by depletion. This allows to reproduce the observed ratio between the (1-0) and the (3-2) line. Our simulations show that varying the thickness of the external layer does not have a great impact on the synthetic spectra, due to the high critical density of \nnh, whose emission arises thus only from the central regions {($n_\mathrm{crit} \gtrsim 10^5 \, \mathrm{cm}^{-3}$, see \citealt{Shirley15})}. Furthermore, N$_2$ is a late-type molecule that can take  $\sim 10$ times longer than CO to form {(e.g. \citealt{Hily-Blant10})}. Thus, the outer less dense envelope may not be rich in N$_2$ enough for \nnh to form efficiently.  The abundance profile for $t = 10^{6}\,$yr and $A_\mathrm{V} = 1$ had to be multiplied by {2.7} to match the observed fluxes. 
 
 \par
\paragraph{\nnd}
Similar considerations to the ones made for the main isotopologue hold also for \nnd. The best fit is found for $t = 10^{6}\,$yr and an external visual extinction of $A_\mathrm{V} = 1\,$mag, even though this last parameter does not affect significantly the simulated spectra (see {Sec. \ref{n2dpAna} for more details}). The models again tend to underestimate the observed fluxes, and we had to increase the abundance profile by a factor of 3.0 to obtain a good agreement.

\paragraph{\hco} In general, the original abundance profiles coming from the chemical network result in a better agreement with the observations of the \hco isotopologues, with respect to the \nnh ones. The best agreement for the main isotopologue is found for the model with $t = 10^{6}\,$yr and an external visual extinction of $A_\mathrm{V} = 5\,$mag, {multiplied by a factor 0.5}. This model presents a high molecular abundance in the external layers of the cores ($X_\mathrm{mol} \gtrsim 10^{-7}$), which is important to correctly reproduce the strong blue-shifted asymmetry of the (1-0) line, due to self-absorption {\citep{Tafalla98}}.

\paragraph{\hcdo} In the chemical model, the abundance profile for the \hcdo is derived from the one of \hco, using the scaling factor valid for the local ISM $\mathrm{^{16}O/^{18}O} = 557$ \citep{Wilson99}. This choice, however, leads to the underestimation of the observed flux. The best fit is found {with the model at} $t = 10^{6}\,$yr and $A_\mathrm{V} = 1\,$mag, multiplied by {6.0}. This late evolutionary stage is important to be able to reproduce the double peak feature seen in the (1-0) spectrum, but we had to decrease the external visual extinction with respect to the main isotopologue. In fact, if the abundance of \hcdo is too high in the external layers of the core, as in the models with $A_\mathrm{V} = 5\,$mag, the resulting synthetic spectra present a blue asymmetry due to the self-absorption, which is not seen in the observations. 

\paragraph{\dco} The three transitions of \dco are best reproduced with the model {at} $t = 10^{6}\,$yr and $A_\mathrm{V} = 1\,$mag. In this case, we decrease by 50\% the original abundance profile to match the observed peak temperatures. \par

\begin{table*}
\centering
\renewcommand{\arraystretch}{1.4}
\caption{{Summary of the models that provide the best fit of each molecule, with the adopted scaling factor for the abundance profile.}  \label{ChemMod}}
\begin{tabular}{cccccc}
\hline
Molecule   &$X_\mathrm{col}$   &  scaling    & $N\mathrm{_{col}^{MOLLIE}}$ \tablefootmark{,a}  & \multicolumn{2}{c}{$T_\mathrm{ex}$ \tablefootmark{b} } \\
             &                                  & factor    &  ($10^{12}\, \rm cm^{-2}$) & line\tablefootmark{c} &  (K) \\
\hline
\multirow{2}{*}{\nnh} & \multirow{2}{*}{$t = 10^6\,$yr, $A_\mathrm{V} = 1$} & \multirow{2}{*}{2.7} & \multirow{2}{*}{17}           & (1-0)               &    5.6  \\
                                     &                                                     &                   &                             & (3-2)               &    5.0   \\
                                     \hline
\multirow{3}{*}{\nnd} & \multirow{3}{*}{$t = 10^6\,$yr, $A_\mathrm{V} = 1$} & \multirow{3}{*}{3.0} & \multirow{3}{*}{$3.5$}           & (1-0)               &     7.7          \\
                                     &                                                     &                   &                             & (2-1)               &     5.0    \\
                                     &                                                     &                   &                             & (3-2)               &   4.8    \\
                                     \hline
\hcdo                 &        $t = 10^6\,$yr, $A_\mathrm{V} = 1$   &    6.0 &     0.18        & (1-0)               &    5.5    \\
\hline
\multirow{3}{*}{\dco} & \multirow{3}{*}{$t = 10^6\,$yr, $A_\mathrm{V} = 1$} & \multirow{3}{*}{0.5} & \multirow{3}{*}{2.5}           & (1-0)               &   7.8  \\
                                     &                                                     &                   &                             & (2-1)               &      5.8   \\
         
                                   &                                                     &                   &                             & (3-2)               &   5.7  \\
\hline
\end{tabular}
\tablefoot{\tablefoottext{a}{{Column density predicted by MOLLIE at the dust peak, computed integrating the molecular abundance, multiplied by the H$_2$ volume density, along the line of sight and convolving to the observation's beam size.}}\\
\tablefoottext{b}{{Excitation temperature values for each line derived as described in the main text and used to derive the column density maps.}}\\
\tablefoottext{c}{{The transitions are labeled with the rotational convention ($J$-$J'$), even when a single hyperfine component is used.}}
}
\end{table*}

\par
\begin{figure*}[h]
\centering
\includegraphics[width =.8 \textwidth]{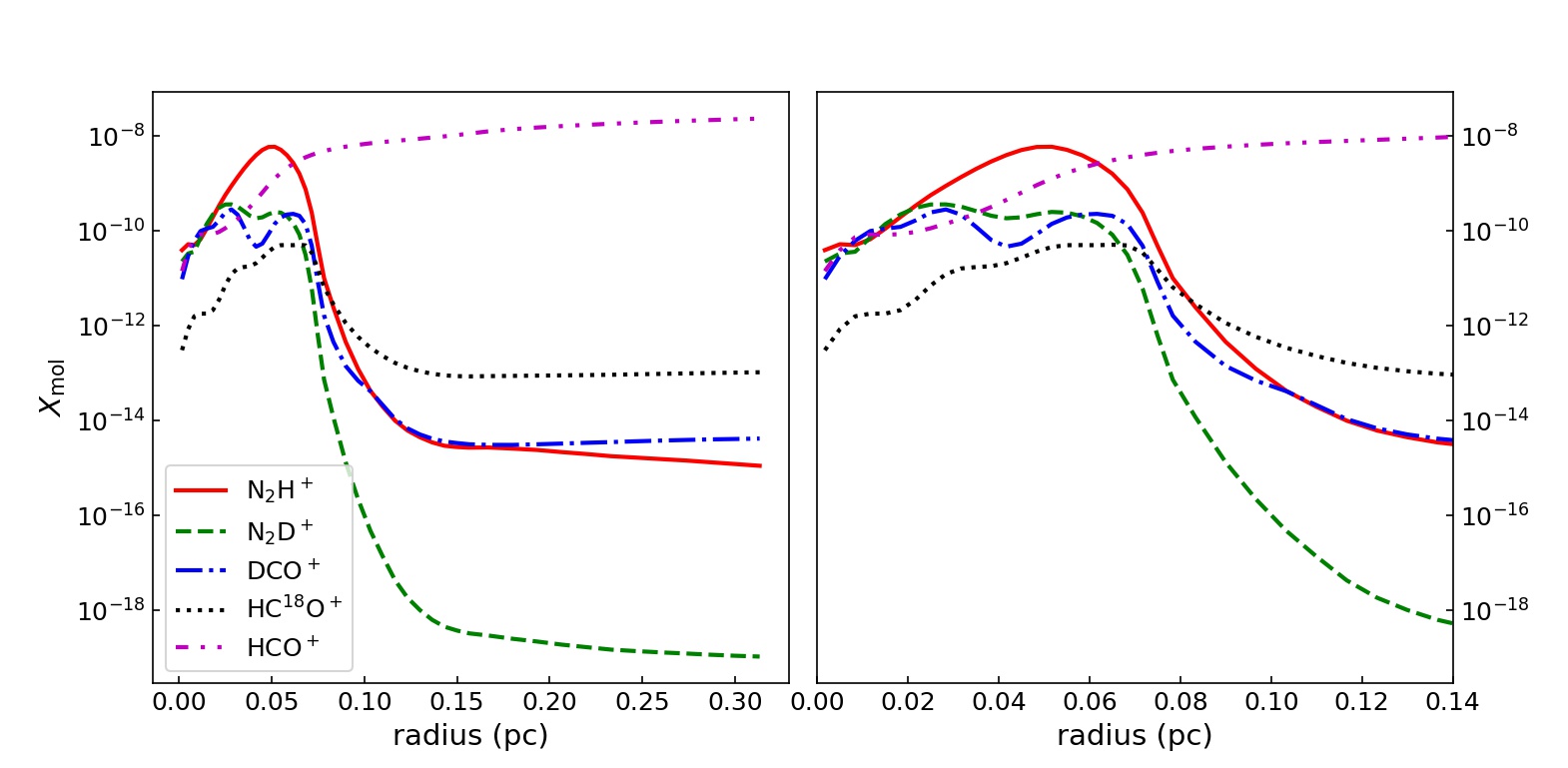}
\caption{\textit{Left panel:} The molecular abundances that provide the best fit to the spectra at the dust peak for the different species, as a function of the core radius. \textit{Right panel:} Zoom-in of the molecular abundances in the inner 0.14$\,$pc. \label{AbProf}}
\end{figure*}

The abundance profiles that provide the best fit solution {for each species} are shown in Figure \ref{AbProf}, after the multiplication for the {corrective} factors. {The adopted models and scaling factors, together with the molecular column density computed by MOLLIE at the core's centre ($N\mathrm{_{col}^{MOLLIE}}$), are summarised in Table \ref{ChemMod}.} It is not surprising that the chemical model results do not provide a perfect match to the observations, and that the various lines cannot be fit with a single model. There are many sources of uncertainty, such as the source model (density and gas/dust temperature structures), chemical modeling parameters and, of course, rate coefficients. Our model does not consider the self-shielding of CO which would affect the \hco abundance in the outer core, for example. A detailed parameter-space exploration of the abundance profiles is beyond the scope of the present paper.

\subsection{   \nnd depletion \label{n2dpAna}}
\begin{figure*}[h]
\centering
\includegraphics[width = 0.5  \textwidth]{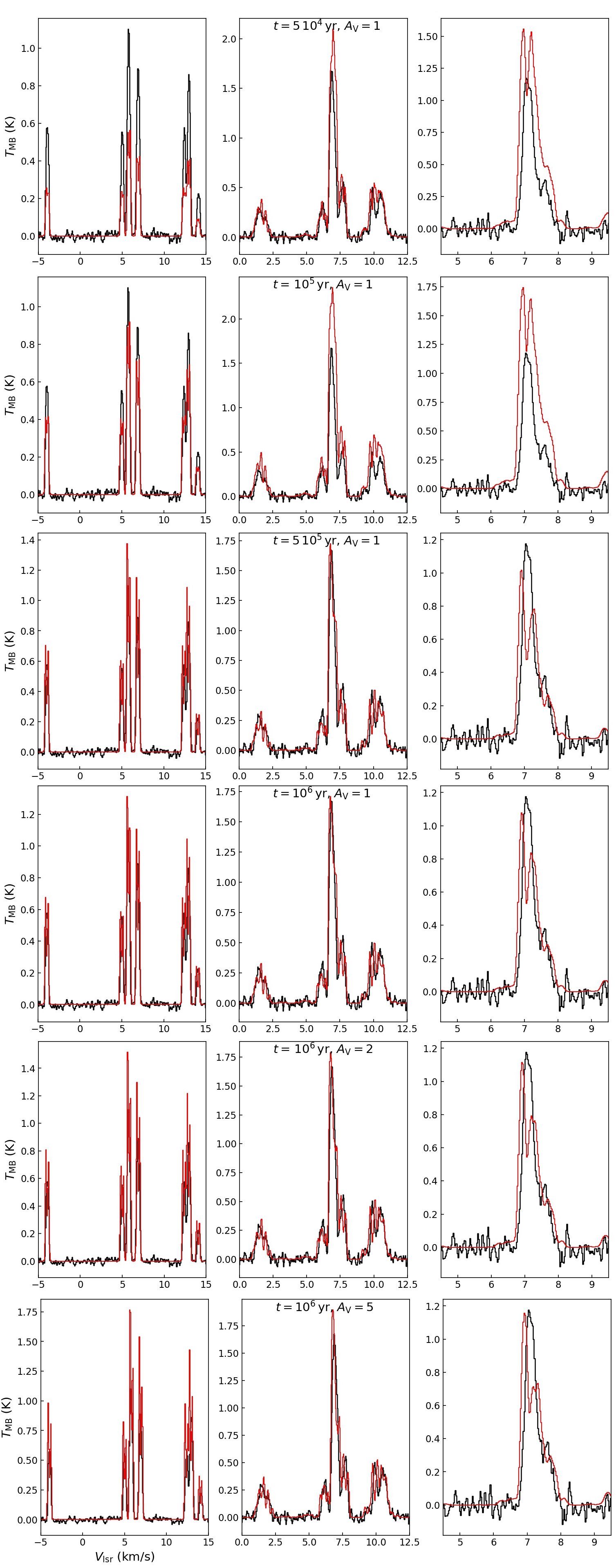}
\caption{  Results of the line radiative transfer for 6 chemical models on the three transitions of \nnd, respectively (1-0), (2-1) and (3-2) from left to right in each panel. The synthetic spectra are shown with red histograms, overlaid to the observations (in black). Each model is labeled with its evolutionary stage and external visual extinction. \label{ndp_allmod}}
\end{figure*}

{
In this Subsection, we analyse in detail the case of \nnd, showing the non-LTE radiative transfer results for 6 out of the 15 abundances profiles that we tested at the dust peak. The abundance profiles are {all multiplied by a factor of 3.0 to allow an easier comparison with the observations.} The obtained spectra, overlaid to the observations, are shown in Figure \ref{ndp_allmod}.}
\par
\begin{figure}[h]
\centering
\includegraphics[width = 0.5  \textwidth]{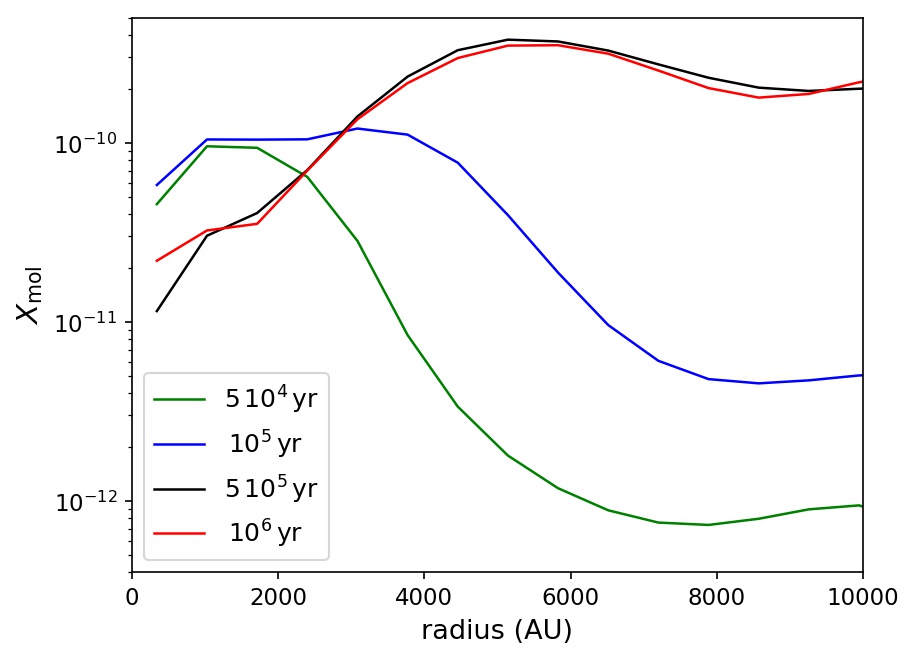}
\caption{  Zoom-in of the central $10000\,$AU of the \nnd abundances predicted by the chemical code for $A_\mathrm{V} = 1\,$mag, at four evolutionary stages:   $t = 5\,10^4\,$yr (green),  $t = 10^5\,$yr (blue),  $t = 5\,10^5\,$yr (black), and  $t = 10^6\,$yr (red). The abundance profiles are multiplied by a factor of 3.0.\label{n2dp_zoom}}
\end{figure}

{
The first four rows from the top in Figure \ref{ndp_allmod} show increasing evolutionary times, keeping the external extinction fixed to $A_\mathrm{V} = 1\,$mag. {The difference between the observations and the model can be quantified by comparing the relative ratio of the peak temperatures of the three lines}. In the observations, this ratio is $T_\mathrm{peak}(1-0) : T_\mathrm{peak}(2-1): T_\mathrm{peak}(3-2) =  1:1.5:1.1$. {In the models with the earlier times ($t = 5\times 10^4$ and $t = 10^5\,$yr), the obtained ratios are $1:3.3:2.5$ and $1:2.6:1.9$, respectively}. This means that these abundance profiles overestimate the (2-1) and (3-2) with respect to the (1-0) line. In the later evolutionary stages ($t = 5\times 10^5$ and $t = 10^6\,$yr), the ratios are {$1:1.2:0.7$ and $1:1.3:0.8$}, closer to the observed one. This trend can be understood by looking at the behaviour of the different abundance profiles in the inner $10000\,$AU, shown in Figure \ref{n2dp_zoom}.  At  earlier times, the \nnd  abundance peaks within the central $2000-3000\,$AU, where the molecular hydrogen density is $n(\mathrm{H_2}) \gtrsim 10^5\, \mathrm{cm}^{-3}$. Since the critical densities of the (1-0), (2-1) and (3-2) transitions are $n_\mathrm{crit} = 7.6 \times10^4\, \mathrm{cm^{-3}}$, $8.7 \times10^5\, \mathrm{cm^{-3}}$, and $ 3.8 \times10^6\, \mathrm{cm^{-3}}$, respectively \citep{Shirley15}, this is also the region where the (2-1) and the (3-2) lines are emitting the most. At later evolutionary times, \nnd  is depleted and the abundance to decreases by more than a factor of five for $r \lesssim 2000\,$AU. {This is due to the fact that its precursor species, N$_2$, starts to freeze-out onto dust grains, and as a consequence the formation rate of \nnd is reduced. The \nnd abundance peak moves outwards to $r \approx 5000\,$AU, where $n(\mathrm{H_2}) \approx 5 \times10^4\, \mathrm{cm^{-3}}$.} Therefore, the (2-1) and (3-2) lines are less excited, while the (1-0) flux keeps rising. This is the first observational confirmation of \nnd partial depletion in cold and dense environments.}

\par
{
We now focus on the influence of the external embedding layer, keeping the evolutionary stage fixed ($t = 10^6\,$yr) and varying the external visual extinction (see the lower three rows of Figure \ref{ndp_allmod}). The synthetic spectra are not very sensitive to this change, both in the flux level and in the spectral profiles, especially concerning the (2-1) and (3-2) lines. This is again due to a combination of molecular abundance and excitation conditions. In fact, increasing $A_\mathrm{V}$ changes significantly the abundance only for $r \gtrsim 10000\,$AU, where the molecular hydrogen density is  $n(\mathrm{H_2}) \lesssim  \,10^4\, \mathrm{cm^{-3}}$. {Therefore, the high frequency transitions are not significantly affected by the change of $A_\mathrm{V}$.}
}

\subsection{Column density and deuteration maps \label{ColDens}}
{In the following, we describe how we derive the column density map of each species, knowing the \ncol value at the core's centre from the non-LTE modelling, via the radiative transfer equations.} The spectra of the {analysed} molecules greatly differ due to their opacity properties and hyperfine structure. For instance, we have optically thin transitions (\hcdo (1-0)), or optically thick ones (\hco (1-0)). Some species present crowded hyperfine structures (such as \nnh and \nnd), while others show a much simpler profile (e.g. \dco, where the hyperfine splitting due to the deuterium spin is not detectable at the given spectral resolution, \citealt{Caselli05}). As a result, we tackle the problem of deriving the column density maps by developing a specific approach for each species individually. \par In all cases, we make use of the well known relations:

\begin{equation}
\label{RadTran}
T_\mathrm{MB} = \eta_{\mathrm{bf}}  \left [ J_\nu (T_\mathrm{ex})   - J_\nu (T_\mathrm{bg})  \right] \left ( 1- e^{-\tau_\nu} \right ) \; ,
\end{equation}
and
\begin{equation}
\label{tau}
\tau_\nu = \sqrt{\frac{\ln 2}{16 \pi^3}}   \frac{c^3 A_\mathrm{ul}  g_\mathrm{u} }{\nu^3 Q(\text{\tex}) \Delta V} e^{- \frac{E_\mathrm{u}}{k_\mathrm{B} \text{\tex}}} \left(  e^{ \frac{h \nu}{k_\mathrm{B} \text{\tex}} } -1 \right) N_\mathrm{col} \; .
\end{equation}
Eq. \eqref{RadTran} describes the main beam temperature observed for a line with a given optical depth $\tau_\nu$ and a given excitation temperature \tex. The function $J_\nu$ is the equivalent Rayleigh-Jeans temperature, and $T_\mathrm{bg} = 2.73\,$K is the cosmic background temperature. $\eta_\mathrm{bf}$ is the beam filling factor, considered to be equal to 1.0 due to the large extension of the core emission compared to the beam size. The second equation relates the optical depth of a given transition to the total column density of the molecule, $N_\mathrm{col}$. The line is considered to be Gaussian, with a full-width-half-maximum equal to $ \Delta V$. $h$ is the Planck constant and $k_\mathrm{B}$ the Boltzmann constant. Eq. \eqref{tau} depends on several spectroscopic constants: the line frequency $\nu$ , the upper state energy $ E_\mathrm{u}$ and degenerancy $ g_\mathrm{u}$, the Einstein coefficient $A_\mathrm{ul}$ and the partition function $Q$. Eq. \eqref{tau} holds both for an entire rotational transition and for an individual hyperfine component, depending on the splitting considered for the levels in the calculation of the partition function and the corresponding level degeneracy used. {We emphasize that Eq. \eqref{RadTran} and \eqref{tau} represent the integrated form of the (differential) radiative transfer equations. This is the appropriate approach since observations yield essentially integrated quantities (such as column density, fluxes, etc.).}

\subsubsection{Column density maps of individual molecules \label{ColDensAna}}
To derive the column density of each molecule, we first smooth the angular resolution of the high-frequency transitions to the one with the largest beam. Then, we fit $N_\mathrm{col}$ pixel by pixel, minimising the residuals between the values of \tmb obtained through Eq. \eqref{tau} and \eqref{RadTran} and the observed peak temperatures for all the available transitions.The observed peak temperatures are obtained with a Gaussian fit to the observed spectra, deriving in this way both \tmb and $\Delta V$ at the same time. When the line presents hyperfine structure, we select a single component. {The choice of the \tex value to be used in Eq. \eqref{RadTran} and \eqref{tau} is made so that this approach at the core's centre agrees with the $N_\mathrm{col}$ value obtained with MOLLIE, within the uncertainties. The \tex values obtained in this way are used to model the whole map. The chosen method to select \tex for each line and the assumption that this quantity is spatially constant are discussed in detail in Appendix \ref{TexDiscuss}.} We now describe in detail on which spectral features we focus to fit $N_\mathrm{col}$ for each species.

\paragraph{\nnh} 
The (1-0) transition of \nnh presents a single, isolated component at the high-frequency end of its spectrum, which is a perfect candidate for our analysis, providing also reliable values for the linewidth. On the contrary, the hyperfine structure of the (3-2) line collapses in a much complicated structure, with a strong shoulder on the red side. For this transition we thus use the peak temperature of the observed spectra, combined with the $\Delta V$ value obtained from the analysis of the (1-0)\footnote{{Several other approaches, such as trying to fit a gaussian avoiding the red-wing component, were tested but proofed to be unsuccessful}.}. For the spectroscopic constants needed in Eq. \ref{tau}, we selected the brightest hyperfine component according to the spectroscopic catalogues (see Appendix \ref{SpectroscopicConstants}). 
\clearpage 

\paragraph{\nnd}
The \nnd (1-0) transition also presents a single, isolated component suitable to our purposes. This is not the case for the other two transitions. For the (2-1) line, we find that the best solution is to consider the brightest feature of the spectra, which is composed by 4 components separated in total by only $\approx 90\,$kHz (see Appendix \ref{SpectroscopicConstants} for details). We fit a single gaussian to this spectral feature in the observations, and we use the additive property of Eq. \eqref{tau}, according to which $\tau_\nu ^\mathrm{tot} = \sum_{i} \tau_\nu ^i$, where $i$ labels the hyperfine components. Finally, in the (3-2) spectra the hyperfine structure is closely blended, so that we consider it a single line\footnote{ We tested this hypothesis at the emission peak by comparing our results with the ones coming from fitting the complete hyperfine structure using GILDAS/CLASS package. We find consistent results within our uncertainties.}.

\paragraph{\hcdo} This is the only molecule for which we have a single transition, and thus we cannot fit $N_\mathrm{col}$ using the procedure described above. However, this transition is optically thin, as confirmed by the radiative transfer performed by MOLLIE, which predicts a maximum value of $\tau_\nu \approx 0.20$ at the dust peak. Therefore, we use the optically thin approximation of Eq. \eqref{RadTran} and \eqref{tau}, following \cite{Caselli02b}:
\begin{equation}
\label{NcolThin}
N_\mathrm{col} = \frac{8 \pi W \nu^3}{c^3 A_\mathrm{ul}} \frac{Q(\text{\tex})}{J_\nu (T_\mathrm{ex})   - J_\nu (T_\mathrm{bg})}    \frac{e^{ \frac{E_\mathrm{u}}{k_\mathrm{B} \text{\tex}}}} {g_\mathrm{u} \left(  e^{ \frac{h \nu}{k_\mathrm{B} \text{\tex}} } -1 \right)} \; ,
\end{equation}
where $W$ is the integrated intensity of the line. We therefore fit a gaussian to the spectrum in each pixel, we compute its integrated intensity and then derive $N_\mathrm{col}$ using Eq. \eqref{NcolThin}.

\paragraph{\dco} The spectra of \dco are the simplest to analyse, since the hyperfine structure due to the deuterium atom is compact if compared to our spectral resolution. We consider the three transitions as single lines, fit a Gaussian profile to each of them and fit the molecular column density minimising the residuals between the synthetic spectra and the Gaussian fits to the observations. \par
\begin{figure*}[h]
\centering
\includegraphics[width = 0.9\textwidth]{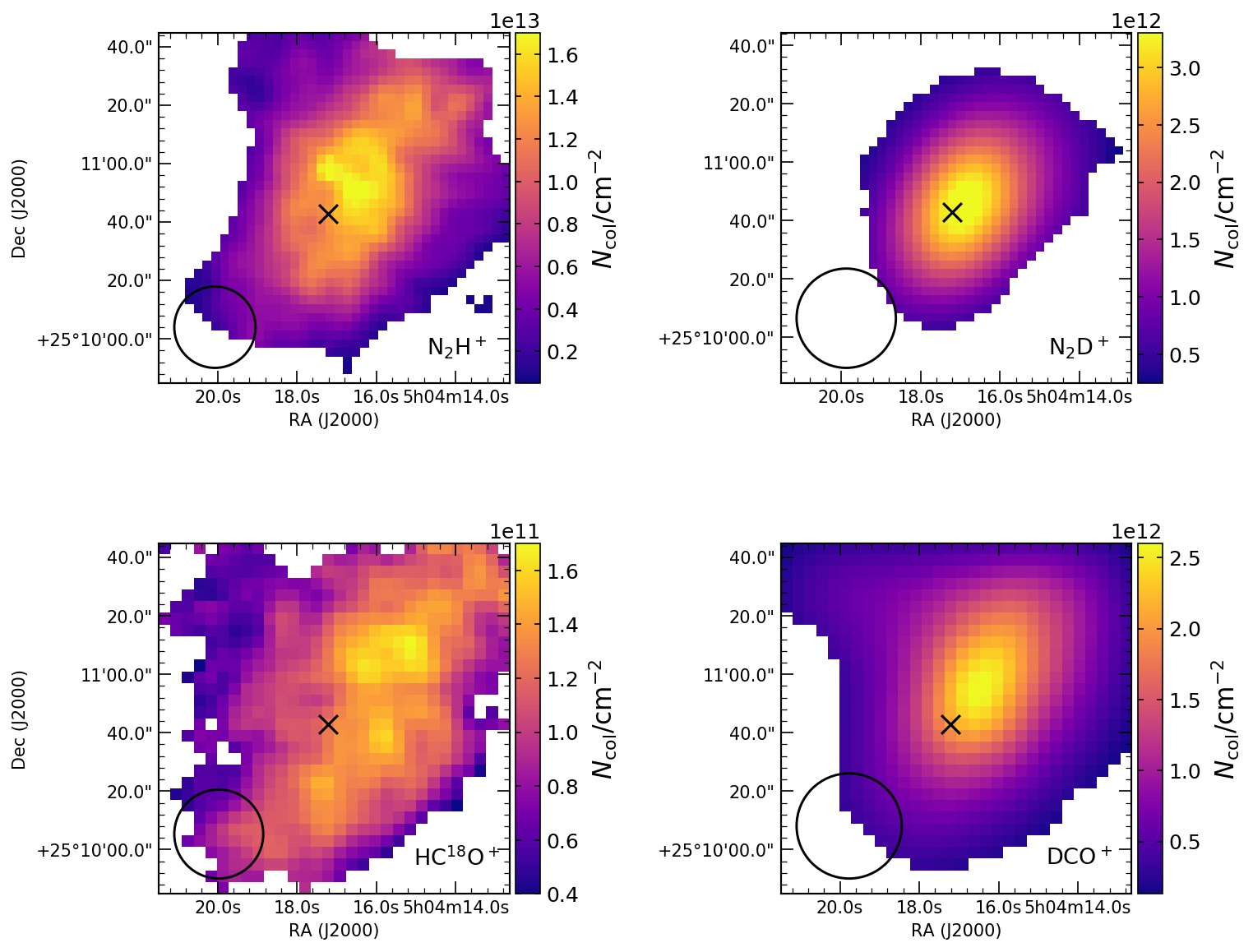}
\caption{The column density maps obtained for each molecule (labeled in the bottom-right corner of each panel). The beam sizes are shown in the bottom-left corners. The black cross represents the position of the millimetre dust peak. \label{ColMaps}}
\end{figure*}

Figure \ref{ColMaps} shows the column density maps obtained for each molecule as just described{, while their peak values are summarised in Table \ref{Ncol_peak}. Our results are consistent with the ones from \cite{Caselli02b}, which are quoted to be accurate within a factor of 2. We however highlight that a direct comparison with other works is difficult (see Sec. \ref{Discuss})}. We report in Appendix \ref{SpectroscopicConstants}  the spectroscopic values used for each transition. {The values of \tex that ensure that the above method provides the same peak column density as computed by MOLLIE are reported in the last column of Table \ref{ChemMod}. They confirm that the assumption of a single \tex value for all transitions is incorrect, since the (3-2) lines present excitation temperatures that are $\approx 2\,$K lower than the respective (1-0) ones.} {For most transitions, \tex is lower than the gas kinetic temperature, suggesting that many of the detected lines are subthermally excited.} \par
One strong assumption of our method is the line Gaussianity. This neglects in fact any spectral asymmetry due for example to the {contraction} motions, which are clearly visible in some spectra. We want to emphasise that any improvement of our analysis is strongly related to the development of a better physical model for L1544. In fact, the one-dimensional model of \cite{Keto15} allows us to solve the radiative transfer problem only at the dust peak, since the fact that L1544 is not spherically symmetric becomes critical moving away from the central point. Since our goal is to investigate the deuteration properties in the whole extended core, we are forced to make further assumptions and simplifications.
\begin{table}[h]
\renewcommand{\arraystretch}{1.5}
\centering
\caption{Peak values of the column densities for the different molecules. \label{Ncol_peak}}
\begin{tabular}{cc}
\hline
Molecule\tablefootmark{a}              & $N_{\mathrm{col}}$/cm$^{-2}$                   \\ \hline
      \nnh               & $ 1.8^{+1.1}_{-0.3} \times10^{13}$                     \\
            \nnd         & $ 3.5^{+2.0} _{-1.0} \times10^{12}$                     \\
            \hcdo   & $ 1.7^{+0.1}_{-0.1} \times 10^{11}$                     \\
\hco &$ 9.5^{+0.6} _{-0.6} \times10^{13}$ \\
       \dco             & $ 2.6^{+1.2} _{-0.8} \times10^{12}$                     \\ \hline
\end{tabular}
\tablefoot{\tablefoottext{a}{The column density of \hco is derived from the one of \hcdo using the standard isotopic ratio $\mathrm{^{16}O/^{18}O} = 557$ (see main text).}}

\end{table}

\begin{figure}[h]
\centering
\includegraphics[width = 0.5\textwidth]{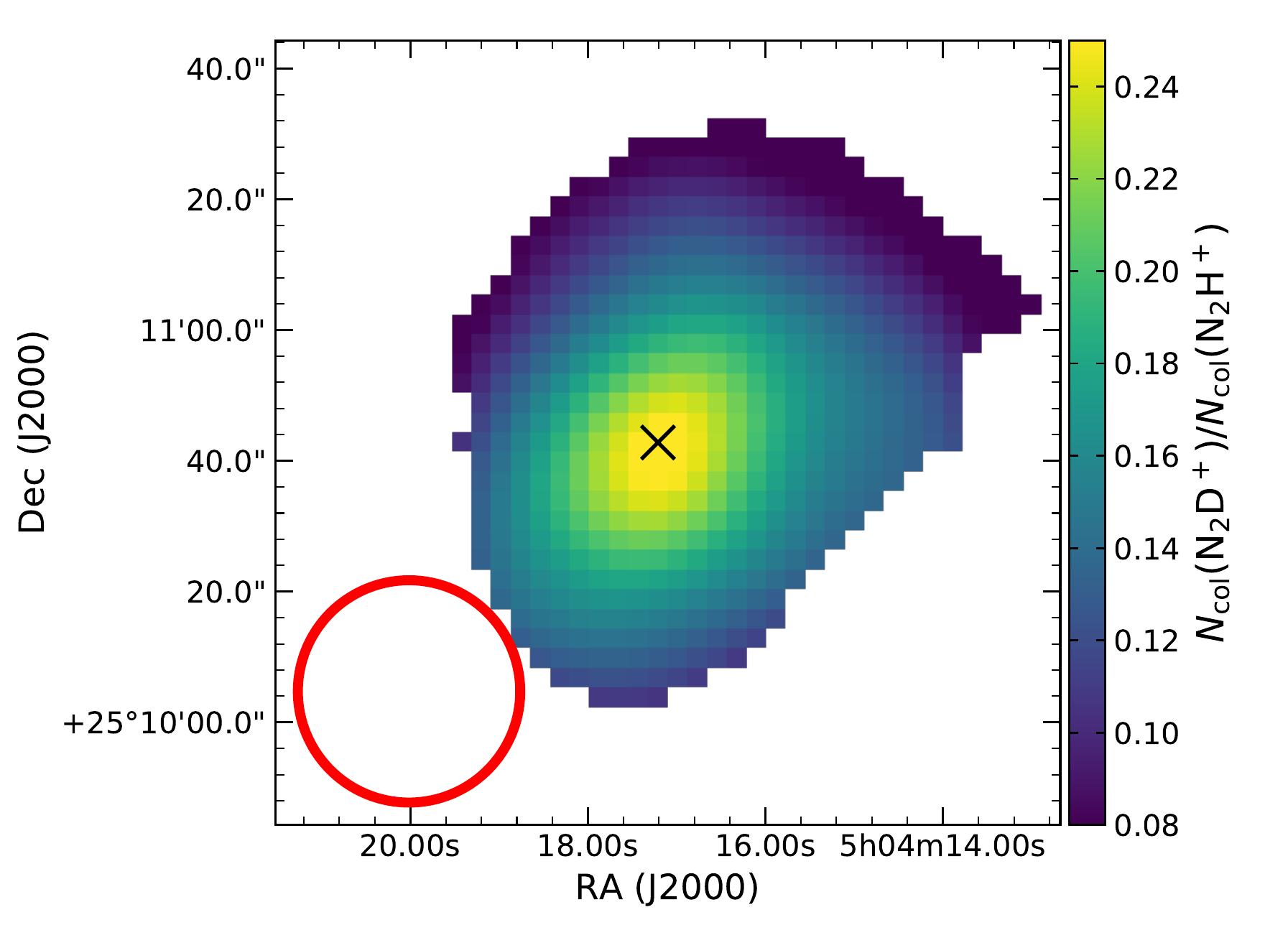}
\caption{D/H ratio obtained for \nnh in L1544. The beam is shown in red, and the black cross represents the dust peak position.\label{deutN2hp}}
\end{figure}

\begin{figure}[h]
\centering
\includegraphics[width = 0.5 \textwidth]{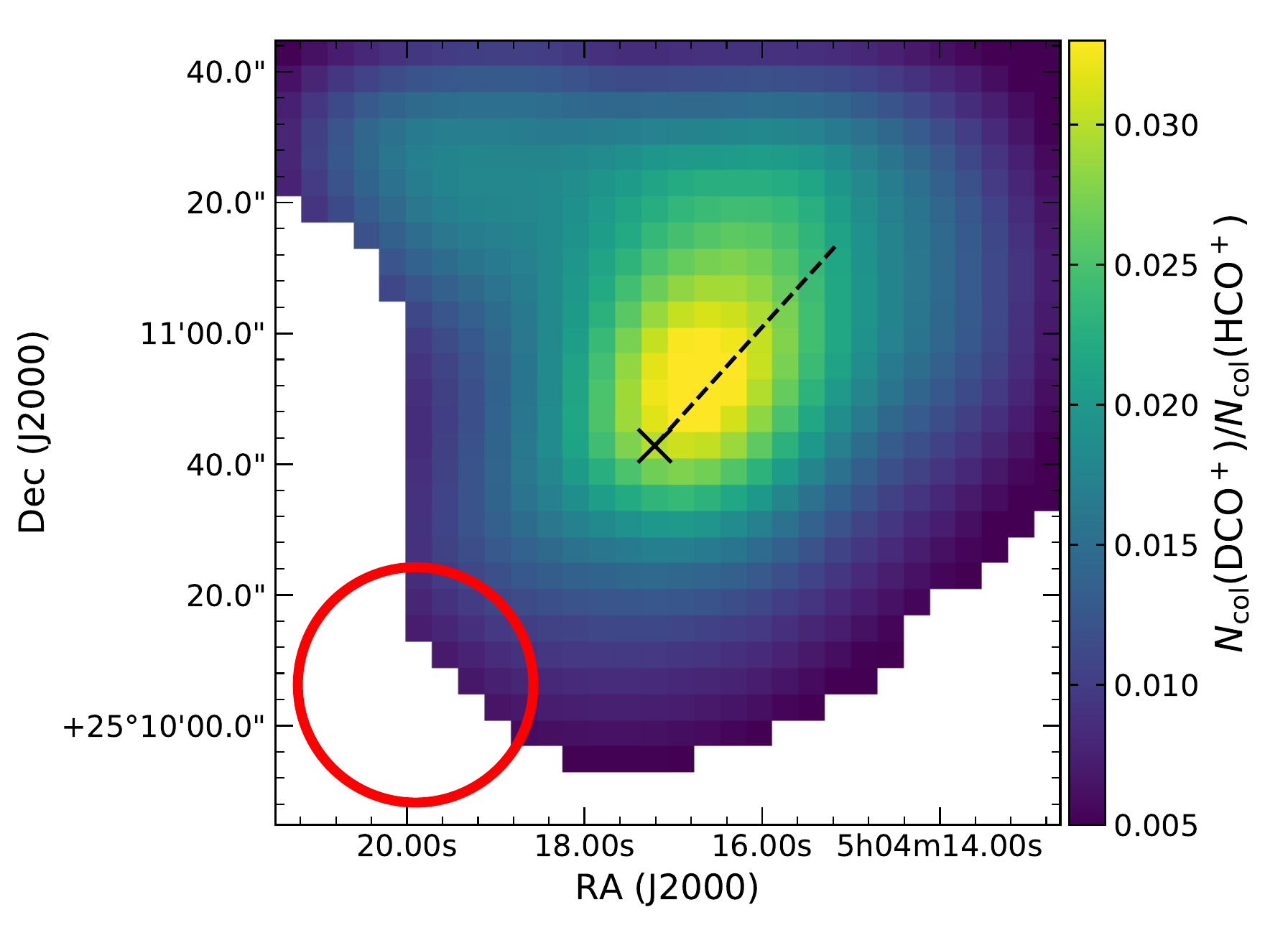}
\caption{Deuterium fraction of \hco obtained in L1544. The beam is shown in red, and the black cross represents the dust peak position. The dashed line is the cut used to produce Figure \ref{trends}. \label{deutHcop}}
\end{figure}

Once the column density maps are ready, we can compute the deuterium fraction, dividing pixel by pixel the column density of the D-bearing species and the one of the main isotopologue. First, the maps' beam sizes are matched to allow a fair comparison. The column density of \hco is derived from the one of \hcdo using the standard isotopic ratio: $N_{\mathrm{col}}({\text{\hco}}) = N_\mathrm{col}(\text{\hcdo})\cdot 557$. Figures \ref{deutN2hp} and \ref{deutHcop} show the resulting images for the D/H ratio.

\subsubsection{{Evaluation of the uncertainties}}
{In the approach adopted to compute the molecular column densities, the main source of uncertainty are the excitation temperature values. In fact, they depend on several other parameters, such as the assumed physical and chemical models. We choose a conservative approach, and decide that the \tex values in Table \ref{ChemMod} are accurate within $0.5\,$K. This choice is supported by the following considerations. First of all, such a variation in \tex translates in an average variation of $20 \sim 30\%$ in the computed column densities, which corresponds to an equal variation of the molecular abundance. We ran again MOLLIE modifying the abundance profiles by $30$\%, and the resulting modeled spectra deviate significantly from the observed ones. This means that our data are indeed sensitive to a variation of $0.5\,$K in \tex. Furthermore, this variation in \tex is the limit suggested by our analysis on the possible spatial variation of this quantity (see Appendix \ref{TexDiscuss}).}\par
{ The column density maps, computed with the \tex values modified by $0.5\,$K, provide thus the searched uncertainties. The uncertainties on the D/H ratios are derived through standard error propagation from the ones on the molecular column density. The median relative error are 41\% for $\mathrm{D/H_{N_2H^+}}$ and 22\% for $\mathrm{D/H_{HCO^+}}$. The lower mean uncertainty on \hco results depends on how sensitive \ncol is to \tex variations, according to Eq. \eqref{tau}. The column density of \nnd, for instance, is affected by a \tex change of $0.5\,$K more heavily than \hcdo, and thus its relative error is higher.}

\section{Discussion \label{Discuss}}
The column density maps presented in Figure \ref{ColMaps} show that different molecules exhibit different morphologies. In particular, one can notice differences in the position of the column density maxima with respect to the dust peak. $N_\mathrm{col}(\text{\nnh})$ and $N_\mathrm{col}(\text{\nnd})$ peak close to the millimetre dust peak, and while the former has a more extended distribution, the latter is highly concentrated in the densest part of the core (as already noted in \citealt{Caselli02b}). The abundance profiles in Figure \ref{AbProf}, despite referring to a one-dimensional core model, confirms these trends. The peak of \nnh abundance is found at a larger radius with respect to \nnd. \par
\begin{figure}[h]
\centering
\includegraphics[width = 0.5 \textwidth]{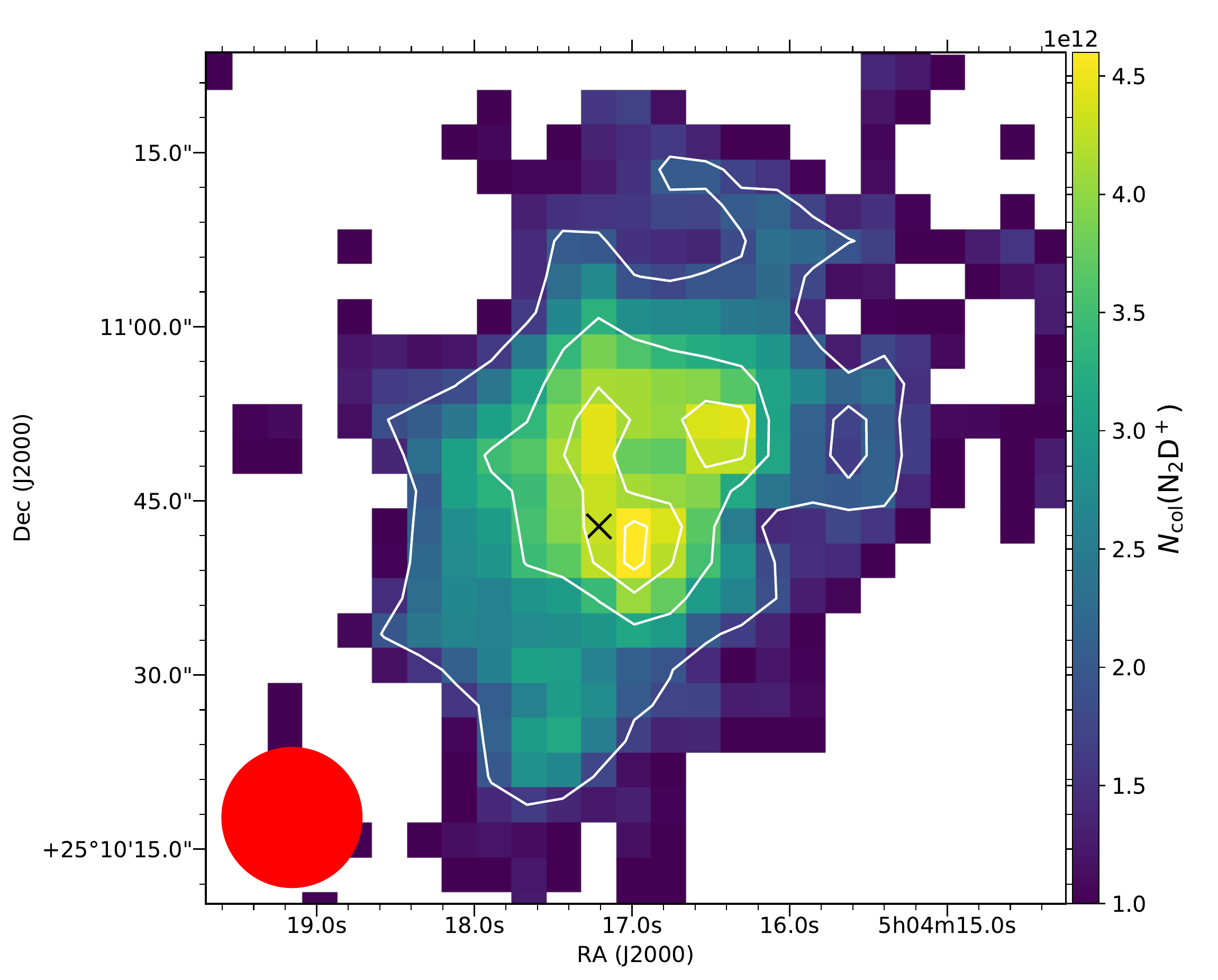}
\caption{  \nnd column density obtained analysing only the (3-2) transition, thus obtaining a spatial resolution almost three times better with respect to the corresponding panel in Fig. \ref{ColMaps}. The white contours represent the integrated intensity of the line, at levels of $[5,9,11,12]\sigma$ ($1\sigma = \rm 0.05 \, K \, km\, s^{-1}$). \label{Nn2dp}}
\end{figure}

{Despite the fact that \nnd is concentrated in the central region, it is interesting to notice that the chemical model that provides the best fit to its transitions predicts some extent of molecular depletion due to freeze-out, {as described in Sec. \ref{n2dpAna}.}} This is to our knowledge the first time that \nnd depletion, which is often predicted by chemical models, is confirmed by observational data. We are aware that our spatial resolution ($35''$ corresponds to $\approx 0.03\,$pc at the distance of L1544) is too poor to investigate this point exhaustively. However, the fact that we need a model with some central depletion to reproduce the observed spectra is a clear confirmation. As a further test, we tried to model all the \nnh and \nnd spectra with MOLLIE using a flat abundance profile with no depletion, which only decreases in the outskirts of the source due to photodissociation. This model is able to reproduce decently the (1-0) lines, but overestimates the fluxes of the higher-J transitions, which have a higher critical density and are thus sensitive to the abundance in the central part of the core. {Furthermore, we derived the molecular column density of \nnd analysing only the (3-2) transition in the optically thin approximation, using Eq. \eqref{NcolThin}. This allows us to obtain a higher resolution map, since the beam size at $\rm 1\, mm$ is $\approx 12''$ (corresponding to $\approx 0.01\, \rm pc$ at L1544 distance). The results are shown in Figure \ref{Nn2dp}. Understandably, the absolute values of \ncol are different from the ones obtain via our complete analysis. However, in this case we are more interested in the morphology of the molecular distribution. Fig. \ref{Nn2dp} shows that \nnd column density is not so concentrated around the core's centre and it does not peak exactly at this position. Instead, it presents multiple secondary peaks, which are significant with respect to our noise level. These features support our hypothesis of an abundance profile which is not increasing or constant in the innermost part of the core, but that on the contrary is decreasing due to the depletion of the molecule from the gas phase.} \par

 Future observational follow-ups with higher resolutions will help us to better understand the extent of \nnh and \nnd depletion, which is of particular interest since both molecules (and \nnd in particular) are expected to preferentially trace the central regions of prestellar cores. Moreover, this provides clues on the freeze-out of the parent molecule N$_2$, which is not observable in cold and dense environments. It is also interesting to mention that ammonia, in comparison, does not exhibit any sign of depletion at the centre of L1544 \citep{Crapsi07}. \par  
\dco morphology is also quite concentrated, similarly to \nnd, but its peak is shifted in the north-west direction. This is confirmed by the integrated intensity maps shown in the bottom panels of Figure \ref{IntInt}. In particular the one for \dco (3-2), which has the best angular resolution ($12''$) shows that the molecular emission peaks $\approx 15 ''$ away from the dust peak.
 This morphology is due to the counter-balance of two chemical mechanisms: (i) the freeze-out onto dust grains of the C- and O-bearing species, which makes them depleted at very high densities and low temperatures \citep{Caselli99, Tafalla02}; (ii) deuterated molecules prefer exactly these physical conditions, as described in the Introduction. This explanation gives also a straightforward interpretation of the $N(\mathrm{HC^{18}O^+})$ map. This molecule presents the highest depletion level of the ones here analysed, as also confirmed by the abundance profile predicted by our chemical network. Its column density peak is found $\approx 35''$ away from the dust peak, in the North-West direction. \par
Another interesting point is that, as described in Sec. \ref{non-ltemod}, we could not reproduce the \hco and \hcdo (1-0) spectra using the same abundance profile, simply scaled by the oxygen isotopic ratio. In fact, to reproduce the strong blue asymmetry shown by the main isotopologue, we needed to adopt a model with a thick envelope, where the molecule is very abundant ($X\mathrm{_{col}(HCO+)} \gtrsim 10^{-7}$). On the contrary, if we use this dense outer layer also for the rare isotopologue, we are not able to correctly fit the spectral profile, since we overestimate the \hcdo abundance in the envelope. We believe that this is a consequence of the selective-photodissociation of \hcdo. While the main isotopologue is so abundant that it is able to self-shield effectively at large radii, the \hcdo abundance is not high enough to block the photodissociating photons that can thus penetrate deeper in the core.  \par
The morphology features previously described for the different column density maps reflect directly on the deuterium fraction maps.  Since the D-bearing isotopologues peak closer to the dust peak than the corresponding main species, the D/H maps exhibit a compact morphology around the dust peak. {The maximum values that we found are $\mathrm{D/H_{N_2H^+}} = 0.26^{+0.15}_{-0.14}$  and $\mathrm{D/H_{HCO^+}} = 0.035^{+0.015}_{-0.012}$}, which are in good agreement with previous estimations (see for instance \citealt{Caselli02b,Crapsi05}). It is however difficult to make a direct comparison with previous works. For example, \cite{Caselli02b} used a constant \tex approach with different excitation temperatures with respect to ours (for instance, $\text{\tex} = 5.0\,$K for \nnh and $4.9\,$K for \nnd). They also lacked some important transitions such as \nnd (1-0) and \dco (1-0), being their frequencies too low for the receivers of the time.
\par
\begin{figure}[h]
\centering
\includegraphics[width = 0.5 \textwidth]{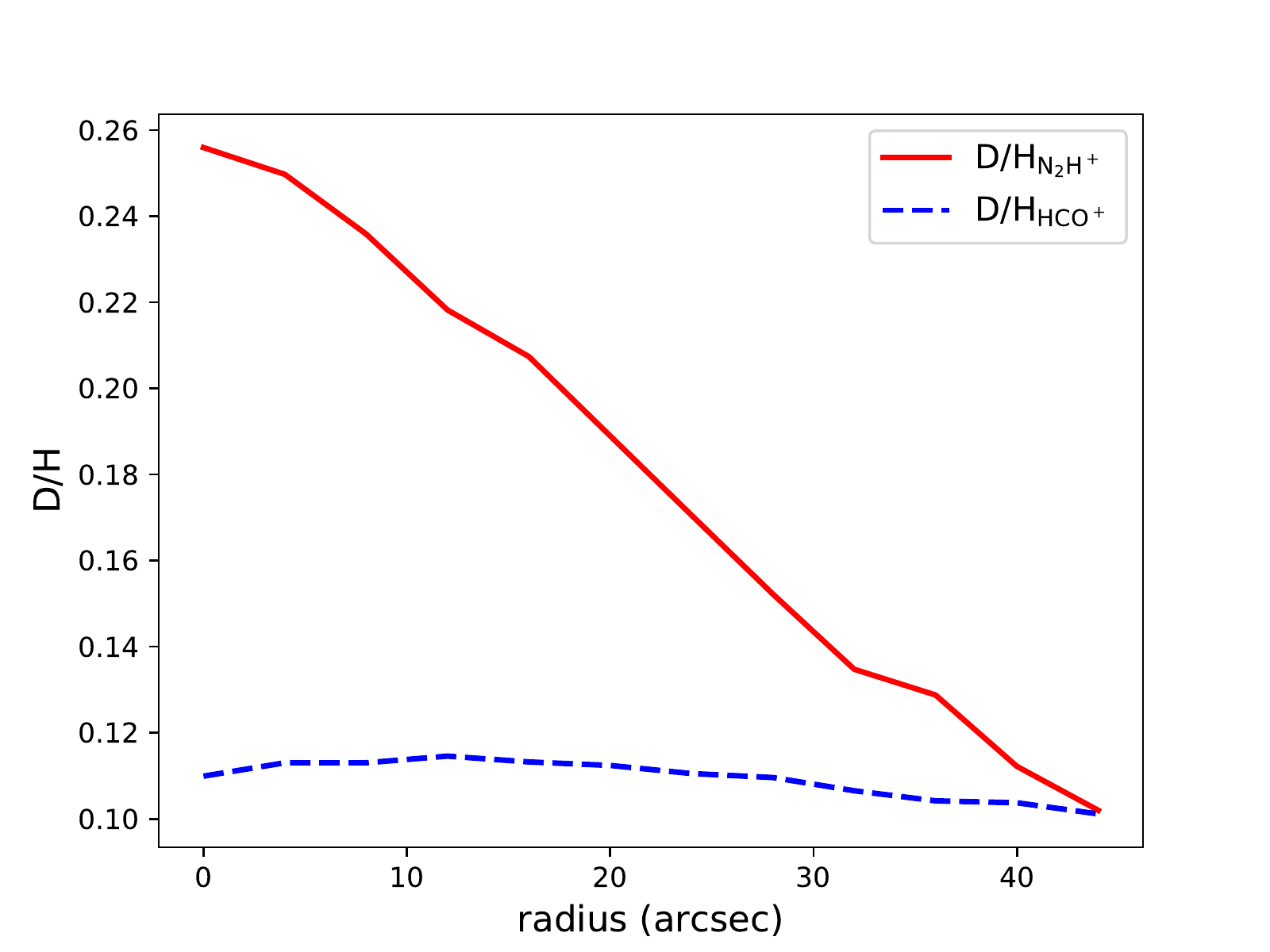}
\caption{Comparison of the trends of $\mathrm{D/H_{N_2H^+}}$ (red) and $\mathrm{D/H_{HCO^+}}$ (blue) along the cut shown in Figure \ref{deutHcop}. The data points of \hco have been shifted upwards by 0.10 to allow an easier comparison. \label{trends}}
\end{figure}

 We find that \nnh is  more deuterated than \hco, as expected being N$_2$ (and \nnh in turn) a late-type molecule {and less affected by depletion \citep{Nguyen18}.} Therefore, it appears in physical condition where CO is already depleted and $\mathrm{H_2D^+}$ is more abundant. {The deuteration level of \nnh decreases from 26\% to 10\% in $45''$, while $\mathrm{D/H_{HCO^+}}$ varies from 3.5\% to 2.0\%. Figure \ref{trends} show the D/H values for the two tracers as a function of radius for a cut crossing the dust peak and the  $\mathrm{D/H_{HCO^+}}$ maximum, shown with a dashed line in Figure \ref{deutHcop}. The deuteration level thus decreases faster for \nnh than for \hco, mainly as a consequence of the fact that \dco is more extended than \nnd with respect to their main isotopologues and that \hco and \dco are more affected by freeze-out in the core centre compared to \nnh and \nnd.}

\section{Conclusions\label{Conclusions}}
In this work we performed a detailed analysis of multiple transitions of \nnh, \nnd, \hcdo, \dco  with recent, high sensitivity observations, which allow us to investigate the molecular properties with high SNR data across the whole L1544 core. Using a non-LTE approach, combined with the molecular abundances computed with chemical models, we derived the excitation conditions of the molecules at the dust peak. Our resulting \tex values confirm that the assumption of a common excitation temperature for different rotational transitions or even for different hyperfine components does not hold for these {molecular ions}. The molecular abundances derived with our chemical models are in general not able to reproduce the observations, especially for \nnh and \nnd for which we had to increase the predicted abundance by up to 300\% to obtain a good fit to the observed spectra. Despite these discrepancies, we are able to draw some important conclusions. \par
All the analysed molecules show some level of depletion in the very central parts of the core, due to freeze-out onto the dust grains. This phenomenon is most prominent for \hco and its isotopologues, but it is seen also for \nnh and to a lesser extent for \nnd. This is of particular interest because it is the first time that \nnd depletion, which is often predicted by chemical models, finds confirmation in observational data. Further, higher-resolution observations will help us investigate this point.  The D/H ratios are peaked very close to dust peak of L1544, and shows different values, being the deuteration of \nnh higher than that of \hco {(26\% and 3.5\%, respectively)}. In an upcoming paper, we will use the results derived here to obtain spatially-resolved information on the ionisation degree, a key parameter in the early stages of star formation.

\begin{acknowledgements}
 
The authors thank the anonymous referee, whose comments helped to increase the quality of the manuscript. E.R. thanks Dr. Jacob Laas for his help in checking the overall language quality. 
\end{acknowledgements}

\bibliography{Literature.bib}

\appendix
\onecolumn
\section{The spectroscopic constants\label{SpectroscopicConstants}}

\begin{figure}[!h]
\centering
\includegraphics[width = 0.55 \textwidth]{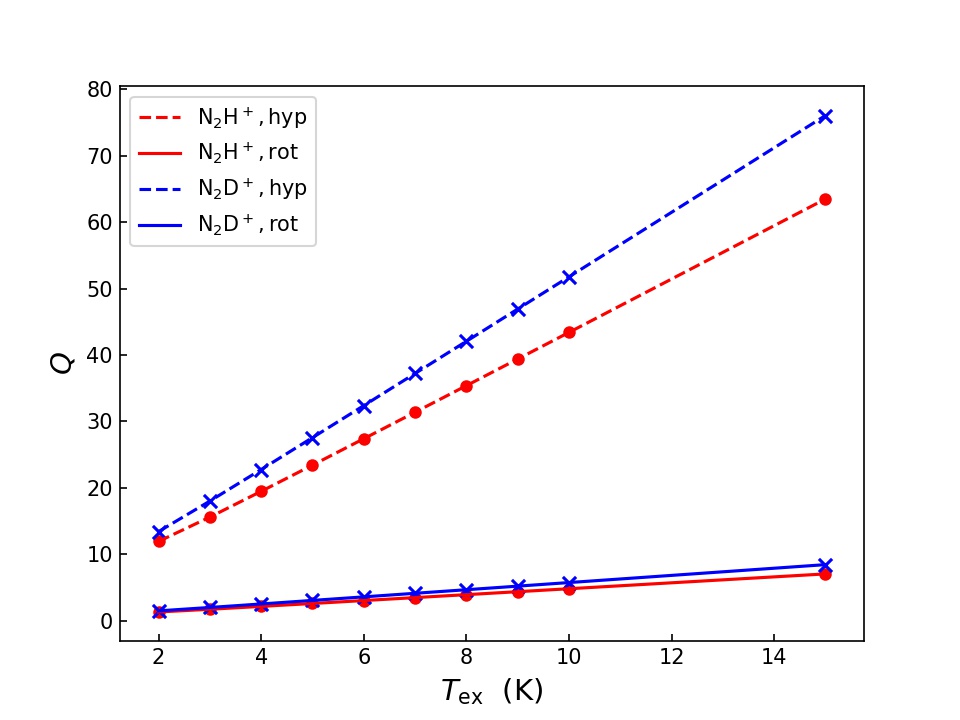}
\caption{The partition function of \nnh (red circles) and \nnd (blue crosses) as a function of excitation temperature. $Q$  for rotational transition, neglecting the hyperfine structure, are indicated with solid lines, while the dashed curves take into account all the hyperfine levels. \label{n2hpQ}}
\end{figure}

\begin{figure}[h]
\centering
\includegraphics[width = 0.55 \textwidth]{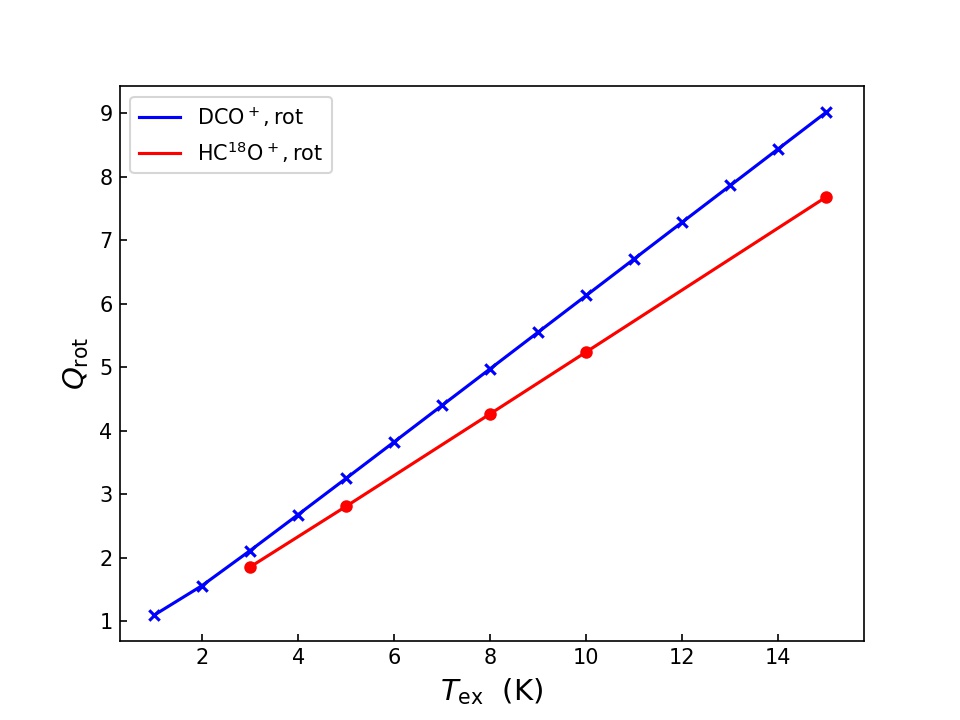}
\caption{Rotational partition function of \dco (blue crosses) and \hcdo (red dots), as a function of \tex. \label{hcopQ}}
\end{figure}

In Table \ref{SpecTable} we report the values of the spectroscopic parameters used to derive the column densities, according to the method explained in Sec. \ref{ColDens}. The references for the shown values are reported in the last column of the Table. Figures \ref{n2hpQ} and \ref{hcopQ} show the partition function for \nnh and \hco (and isotopologues), respectively. In the considered range of temperatures, the correlation is almost linear. 

\begin{table}[!h]
\renewcommand{\arraystretch}{1.3}
\centering
\caption{Spectroscopic values for the transitions used in the analysis to derive the molecular column densities.\label{SpecTable}}

\begin{tabular}{ccr@{.}lcr@{.}lr@{.}lc}
\hline
Rot. Transition & Hyperfine comp.\tablefootmark{a}& \multicolumn{2}{c}{$\nu$ (GHz)}   & $g_\mathrm{u}$ & \multicolumn{2}{c}{$E_\mathrm{u}/k_\mathrm{B}$ (K)} & \multicolumn{2}{c}{$A_\mathrm{ul}/10^{-5}$ ($\text{s}^{-1}$)} & Notes\tablefootmark{c} \\ \hline
\multicolumn{10}{c}{\nnh}                                                                                                              \\
$J=1-0$             &    $J,F_1,F = 1,0,1 \leftarrow 0,1,*$\tablefootmark{b}          & 93&1762604  & 3              & 4&472               & 3&628          &   1      \\
$J=3-2 $            &    $J,F_1,F = 3,4,5 \leftarrow 2,3,4$               & 279&5118572 & 11             & 13&41     & 1&259$\,10^2$           &   1    \\ \hline
\multicolumn{10}{c}{\nnd}                                                                                                              \\
$J=1-0 $            &      $J,F_1,F = 1,0,1 \leftarrow 0,1,*$     & 77&1121207  & 3              & 3&701               & 2&057    &    2    \\
$J=2-1$             &       $J,F_1,F = 2,3,3 \leftarrow 1,2,2$   & 154&2170958 & 7              & 11&10            & 18&08      &  2     \\
$J=2-1  $           &       $J,F_1,F = 2,3,2 \leftarrow 1,2,1$& 154&2171220 & 5              & 11&10       & 15&63           &    2   \\
$J=2-1  $           &     $J,F_1,F = 2,2,3 \leftarrow 1,1,2$     & 154&2171273 & 7              & 11&10     & 14&17           &  2    \\
$J=2-1  $           &  $J,F_1,F = 2,3,4 \leftarrow 1,2,3$ & 154&2171888 & 9              & 11&10       & 19&74         &   2        \\
$J=3-2  $           & -               & 231&3218611 & 7              & 22&20       & 71&38             &  2 \\ \hline
\multicolumn{10}{c}{\dco}                                                                                                              \\
$J=1-0 $            & -               & 72&0393123  & 3              & 3&457      & 2&206          &   3  \\
$J=2-1 $            & -               & 144&0772854 & 5              & 10&37  & 21&18        &    3      \\
$J=3-2 $            & -               & 216&1125804 & 7              & 20&74      & 76&58           &  3  \\ \hline
\multicolumn{10}{c}{\hcdo}                                                                                                             \\
$J=1-0  $           & -               & 85&1622231  & 3              & 4&087       & 3&645         &     4    \\ \hline
\end{tabular}
\tablefoot{
\tablefoottext{a}{When no hyperfine component is indicated, the transition is intended as single transition.}\\
\tablefoottext{b}{The notation $J,F_1,F \leftarrow J',F_1',*$ indicates that there are multiple but degenerate lower states, which have the same energy but different $F$ quantum number.}
\tablefoottext{c}{The spectroscopic data are taken from: 1) Our calculation based on data from \citep{Cazzoli12}. 2) Our calculation based on data from \citep{Dore04, Amano05, Yu15}. 3) Our calculation based on data from \citep{Caselli05, Lattanzi07}. 4) From Bizzocchi et al. (in prep.). }
}
\end{table}

\section{  Molecular abundance profiles \label{CompleteModels}}
\begin{figure*}[h]
\centering
\includegraphics[width = 0.85 \textwidth]{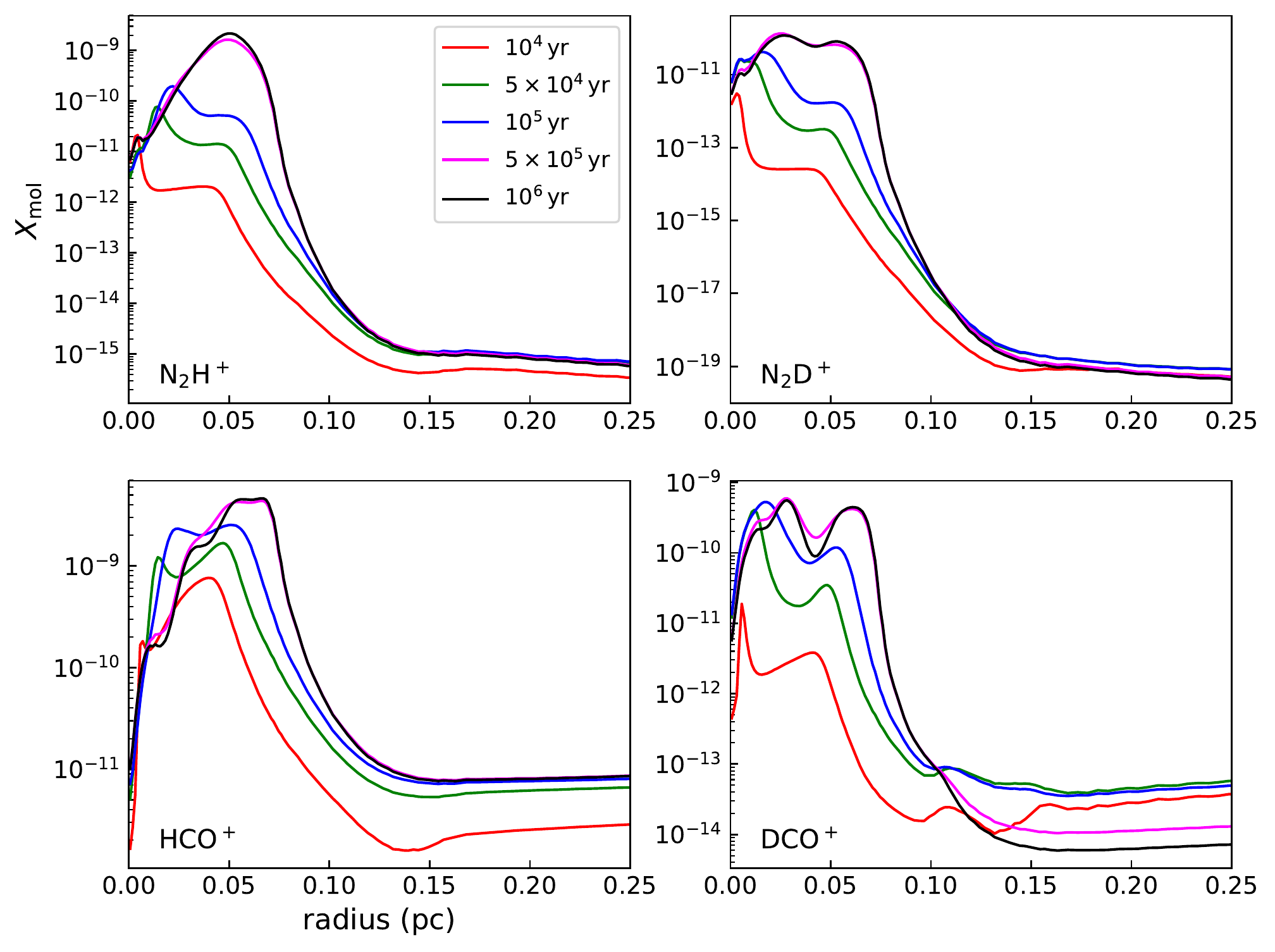}
\caption{Molecular abundances in the model {with $A_\mathrm{V} = 1\,$mag} for \nnh (top-left), \nnd (top-right), \hco (bottom-left), and \dco (bottom-right). The colors represent the {different time-steps: $10^{4}\, $ yr (red), $ 5\times 10^{4}\, $ yr (green), $10^{5}\,$  yr (blue), $5 \times10^{5}\, $ yr (purple), and $10^{6}\, $ yr (black)}.  \label{ChemMods1}}
\end{figure*}

\begin{figure*}[h]
\centering
\includegraphics[width = 0.85 \textwidth]{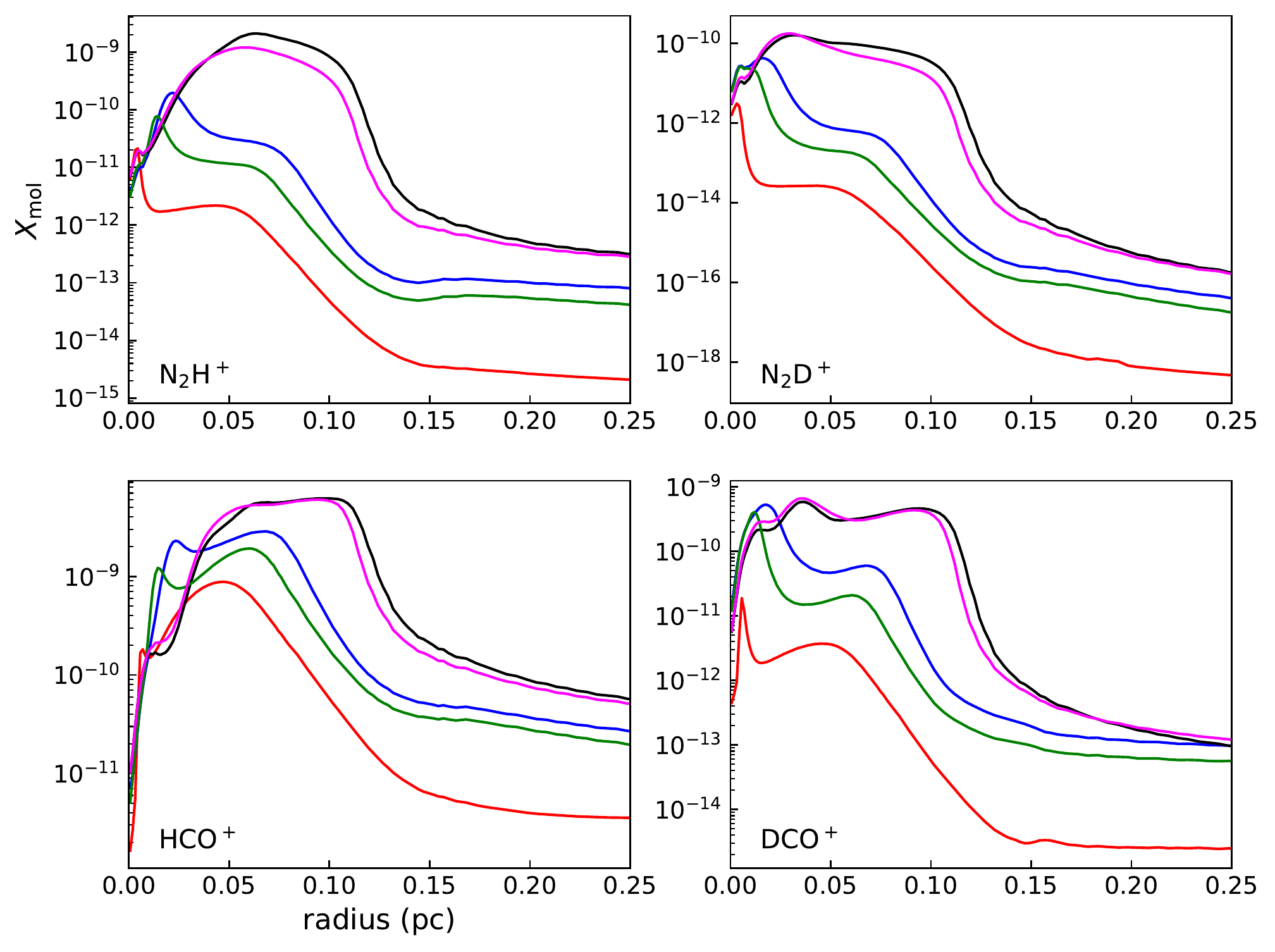}
\caption{  Same as Fig. \ref{ChemMods1}, for $A_\mathrm{V} =2\,$mag. \label{ChemMods2}}
\end{figure*}

\begin{figure*}[h]
\centering
\includegraphics[width = 0.85 \textwidth]{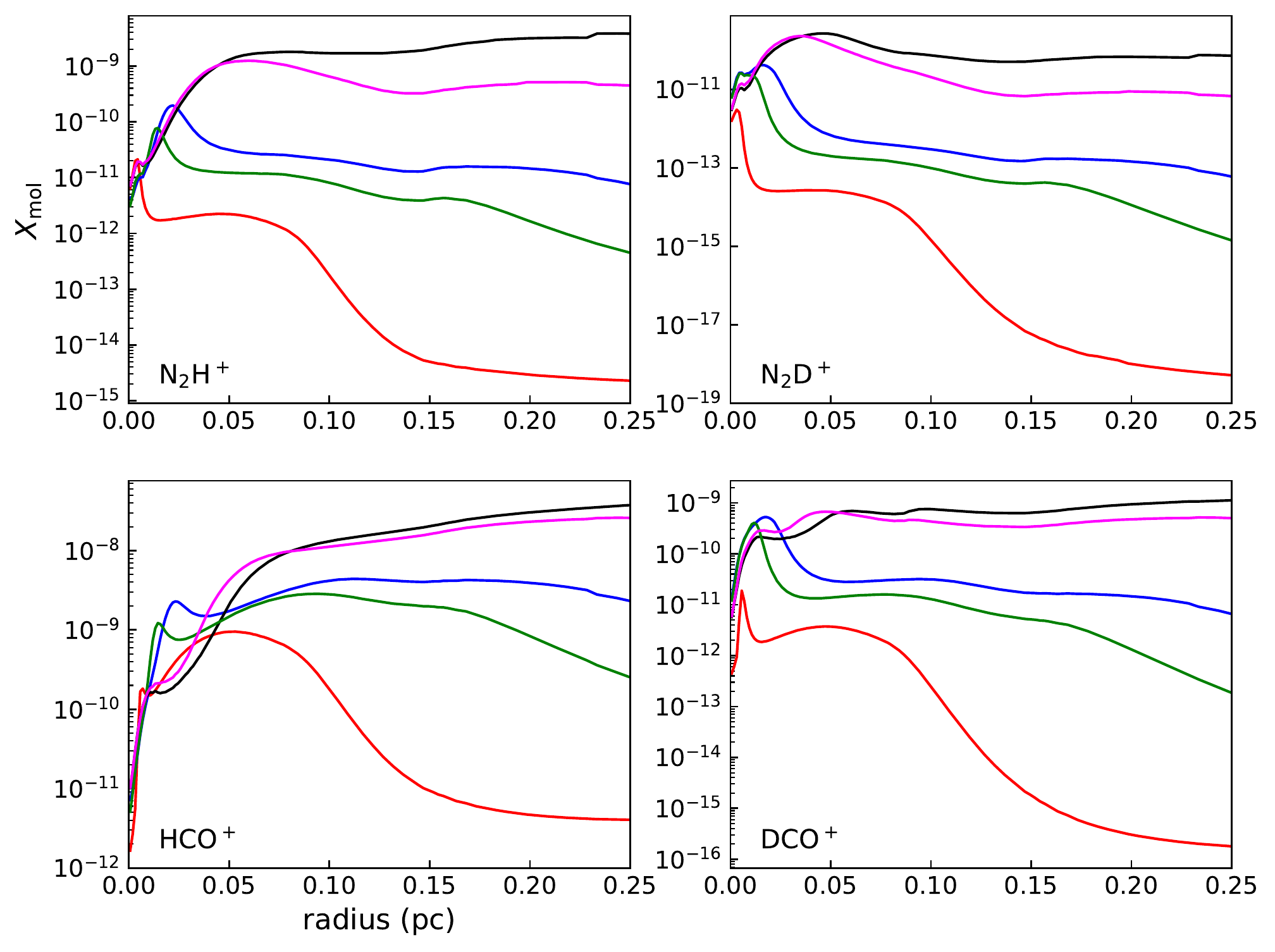}
\caption{  Same as Fig. \ref{ChemMods1}, for $A_\mathrm{V} =5\,$mag. \label{ChemMods3}}
\end{figure*}

{In this Appendix, we plot the abundance profiles computed with our chemical network and tested with the non-LTE approach. Fig. \ref{ChemMods1},  \ref{ChemMods2}, and \ref{ChemMods3} show the models at the 5 different time-steps, respectively for $A_\mathrm{V} = 1, 2, \text{ and }5$.}


\section{Discussion on the excitation temperature\label{TexDiscuss}}
As described in Sec. \ref{Analysis}, a main issue to derive the molecular column density when its rotational lines are observed consists in determining the excitation temperatures of the transitions. When \tex is not the same for all the lines (or when just one line is available), the problem becomes degenerate. A non-LTE analysis is thus necessary to constrain these parameters (\ncol and \tex) independently. \par
One could argue that MOLLIE can be used to derive directly the \tex values. This is only partially true. MOLLIE, as any non-LTE radiative transfer code that performs ray tracing, does not operate directly with the excitation temperature quantity. It propagates and counts photons, and it solves the statistical equilibrium equations for the energy level populations. The latter are indeed related to \tex through the Boltzmann equation. However, the \tex obtained in this way are only local values, since they can be computed cell by cell\footnote{In our simulation, the finest spatial grid has a resolution of $3\,10^{-3}\,$pc, orders of magnitude smaller than the resolution of the observed spectra.}. The results is thus a one-dimensional radial profile of \tex for each transition. \par
On the other hand, the \tex that appears in the equations of radiative transfer is an average quantity, integrated along the line of sight and over the beam size, which relates the observed fluxes with the molecular column density for the beam of that specific observation. It is not straightforward to link the local \tex profiles obtained in MOLLIE with the \tex values needed to compute \ncol, especially because \tex is an intensive quantity. An approach that we have tested is to integrate the \tex profiles along the line of sight, using the H$_2$ column density as a weighting function. Since H$_2$ is the main collisional partner, this choice seems reasonable. However, it neglects other important parameters that affect \tex, such as the gas temperature and the density of the molecule itself. Another point that must be taken into account is that, contrary to \tex, the molecular abundance is an extensive quantity, that can be multiplied by $n \rm (H_2)$ and integrated along the line-of-sight to derive the molecular column density without any ambiguity. It is a natural requirement that the two methods to derive \ncol, i.e. the physical and chemical modelling done with MOLLIE and the solution of Eq. \eqref{RadTran} and \eqref{tau}, must be consistent and provide the same result at the core's centre. With this constrain, \tex becomes a free parameter that can be tuned to ensure the required consistency. This is the selected method to derive the \tex values reported in Table \ref{ChemMod}. \par
We remark that this approach is possible only at the dust peak, because the physical and chemical models are spherically symmetric and cannot reproduce the core's asymmetries that the observations unveils. Furthermore, the model of \cite{Keto15} has been specifically developed to reproduce the dust peak. 
\subsection{\tex spatial variations}
A strong assumption that we make in Sec. \ref{ColDens} is that the \tex values found as just described are constant across all our map coverage. This can be objected, since one can expect that molecules are less excited in the outskirts of the core, where the gas is less dense. In order to investigate this point, we tried to derive the \dco column density using a decreasing \tex profile for each line. \dco is a representative case, because it has a more extended distribution than for instance \nnd and three transitions are available at the same time. We realised that a decrease of more than $0.5-1.0\,$K within $\approx 60''$ (the limit of our maps' coverage) is enough to make our \ncol fitting routine to break down. With these profiles, the resulting \ncol maps present unphysical features, such as values at the core's border several times higher that at the integrated intensity peak, and/or unrealistic morphologies. \par
\begin{figure*}
\centering
\includegraphics[width = 0.85 \textwidth]{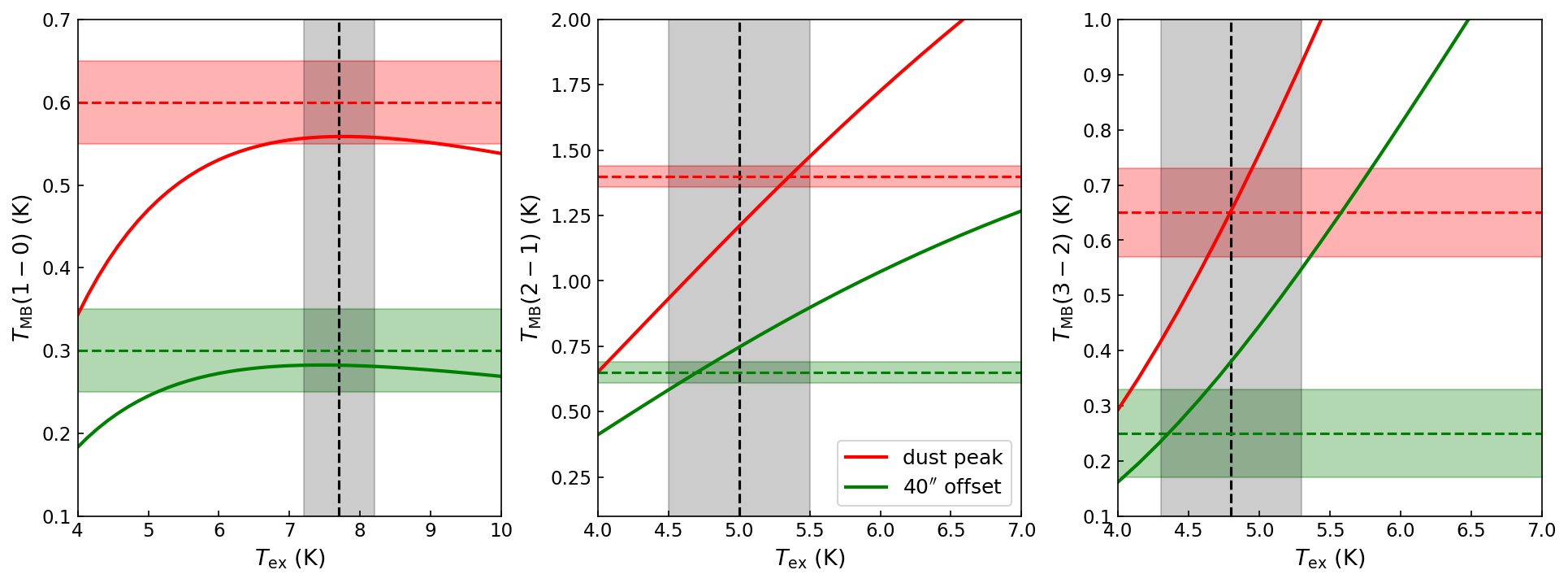}
\caption{ In each panel, the horizontal dashed lines are the observed peak temperatures of the selected \nnd transition at the dust peak (red), and at 40'' of offset (green). The shadowed areas represent observational uncertainties. The solid curves indicate \tmb as a function of \tex, obtained via the radiative transfer equations and using the column density values predicted by MOLLIE at the two offsets. The vertical dashed line is the \tex value used in the analysis of the maps, and the grey shade is its uncertainty ($0.5\,$K). The panels refer to the isolated component of \nnd (1-0) (left), the central component of \nnd (2-1) (centre), and the \nnd (3-2) line (right). \label{n2dp_tex}}
\end{figure*}

In order to further test this assumption, we focus on \nnd. As already mentioned, MOLLIE and our physical and chemical models are not able to reproduce the observed spectra outside the dust peak. However, \nnd is sufficiently centrally concentrated {and symmetric around the core's centre} to allow a decent fit to the observations also outside {dust peak}. {In particular, we were able to compare the synthetic spectra extracted from the model at $40''$ offset with the observed signals obtained by averaging the data in a ring at a distance of $40''$ from the core's centre. We can therefore} constrain \ncol at two different offsets, and verify that the \tex values that reproduce these column densities are equal {within the uncertainties}. Figure \ref{n2dp_tex} shows the results of this analysis. The red and green dashed lines represent the \tmb observed values of the line components used to derive \ncol (see Sec. \ref{ColDensAna}), respectively at the dust peak and at $40''$ offset. The shaded area indicate the spectra rms\footnote{ These uncertainties are lower limits, since they do not take into account the calibration uncertainties (up to 15\%).}. The solid curves indicate the \tmb of each line as a function of \tex, obtained using Eq. \eqref{RadTran} and \eqref{tau} using the column density values predicted by MOLLIE at the two offsets ($3.5$ and $1.6 \times 10^{12} \, \rm cm^{-2}$, respectively). Finally, the black dashed lines show the \tex values adopted in this work, and the grey shadows indicate a $0.5\,$K variation of this quantity. All the observations are smoothed at the same beam size ($34''$), which is the resolution of the final \ncol map. As one can notice, for all three transitions the solid lines cross the interceptions of the shaded areas, {suggesting} that within our uncertainties we are not able to detect a significant change of \tex. \par
We want to highlight that our analysis does not imply that \tex is indeed constant across the whole core, but only that our observations are not sensitive to its variation and that they can be analysed assuming a constant value. Most likely, this is also due to the small coverage of our maps, which are $2'\times 2'$ in size, meaning that in a core radius we have at most $2-3$ independent beams. All this said, we assume that our \tex values reported in Table \ref{ChemMod} are accurate within $0.5\,$K.

\end{document}